\documentclass[aip, amsmath,amssymb,reprint,]{revtex4-1}

\usepackage{graphicx}
\usepackage{dcolumn}
\usepackage{bm}

\usepackage[utf8]{inputenc}
\usepackage[T1]{fontenc}
\usepackage{mathptmx}
\usepackage{color}
\usepackage{graphicx}
\usepackage{tikz,pgf}
\usepackage{array}
\usepackage{xcolor}
\usepackage[dvips]{epsfig}
\usepackage{epsfig}
\usepackage{enumerate}
\usepackage{bbold}
\usepackage{microtype}

\usepackage{soul}
\usepackage[normalem]{ulem}

\usepackage{simplewick}

\newcommand{\ga}{\ga}

\newcommand{\bl}{\begin{linenomath*}}
\newcommand{\el}{\end{linenomath*}}

\newcommand{\bea}{\begin{eqnarray}}
\newcommand{\eea}{\end{eqnarray}}

\renewcommand{\ga}{\hat\gamma}

\definecolor{dgreen}{rgb}{0.0, 0.5, 0.0}

\usepackage{hyperref}
\graphicspath{{figures/}}

\begin{document}

\preprint{AIP/123-QED}

\title[]{Intermolecular forces and correlations mediated by a phonon bath}

\author{Xiang Li}
\affiliation{IST Austria (Institute of Science and Technology Austria), Am Campus~1, 3400 Klosterneuburg, Austria}

\author{Enderalp Yakaboylu}
\affiliation{IST Austria (Institute of Science and Technology Austria), Am Campus~1, 3400 Klosterneuburg, Austria}

\author{Giacomo Bighin}
\affiliation{IST Austria (Institute of Science and Technology Austria), Am Campus~1, 3400 Klosterneuburg, Austria}

\author{Richard Schmidt}
\affiliation{Max Planck Institute of Quantum Optics, Hans-Kopfermann-Str.~1, 85748 Garching, Germany}
\affiliation{Munich Center for Quantum Science and Technology (MCQST), Schellingstr.~4, 80799 M\"unchen, Germany}

\author{Mikhail Lemeshko}
\affiliation{IST Austria (Institute of Science and Technology Austria), Am Campus~1, 3400 Klosterneuburg, Austria}

\author{Andreas Deuchert}
\affiliation{IST Austria (Institute of Science and Technology Austria), Am Campus~1, 3400 Klosterneuburg, Austria}
\affiliation{Institute of Mathematics, University of Zurich, Winterthurerstrasse 190, 8057 Zurich, Switzerland}

\date{\today}

\begin{abstract}


Inspired by the possibility to experimentally manipulate and enhance chemical reactivity in helium nanodroplets, we investigate the effective interaction and the resulting correlations between two diatomic molecules immersed in a bath of bosons. By analogy with the bipolaron, we introduce the \emph{biangulon} quasiparticle describing two rotating molecules that align with respect to each other due to the effective attractive interaction mediated by the excitations of the bath. We study this system in different parameter regimes and apply several theoretical approaches to describe its properties. Using a Born-Oppenheimer approximation, we investigate the dependence of the effective intermolecular interaction on the rotational state of the two molecules. In the strong-coupling regime, a product-state ansatz shows that the molecules tend to have a strong alignment in the ground state. To investigate the system in the weak-coupling regime, we apply a one-phonon excitation variational ansatz, which allows us to access the energy spectrum. In comparison to the angulon quasiparticle, the biangulon shows shifted angulon instabilities and an additional spectral instability, where resonant angular momentum transfer between the molecules and the bath takes place. These features are proposed as an experimentally observable signature for the formation of the biangulon quasiparticle. Finally, by using products of single angulon and bare impurity wave functions as basis states, we introduce a diagonalization scheme that allows us to describe the transition from two separated angulons to a biangulon as a function of the distance between the two molecules.


\end{abstract}

\maketitle

\section{\label{sec:level1}Introduction}

The emergence of effective, bath-mediated interactions between quantum particles is an important phenomenon with many examples from different subfields of physics. One of the most prominent is the effective phonon-mediated interaction between two polarons~\cite{Devreese15}, that is, between two electrons in a crystal that are dressed by a cloud of lattice excitations. This effective attractive interaction can overcompensate the Coulombic repulsion between the electrons and results in the formation of the bipolaron quasiparticle\cite{devreese2009frohlich,kashirina2010large} -- a bound state that has been proposed as one of the mechanisms behind high-temperature anomalous superconductivity \cite{alexandrov2003theory}. In case of sufficiently strong electron-phonon interactions, also more complex polaronic structures such as electronic Wigner crystals \cite{quemerais1998polaron,fratini2002polarization,iadonisi2007formation}, polaron molecules and clusters \cite{kusmartsev2001electronic,perroni2004formation,bruderer2007polaron} can form. Moreover, the electron-phonon coupling has been used to explain the thermodynamic and optical properties of quantum dot devices~\cite{fomin1998photoluminescence,klimin2004ground}. Finally, attractive electron interactions mediated by phonons are found to be able to overcome the direct Coulomb repulsion in deformable molecular quantum dots, paving the way for the realisation of polaronic memory resistors \cite{alexandrov2003memory,alexandrov2009polaronic}.

In the context of ultracold atoms various theoretical methods have been developed to study bath-mediated correlations in Bose-Einstein condensates in the case of attractive/repulsive couplings \cite{jorgensen2016observation,hu2016bose} and for weakly\cite{casteels2011many}/strongly interacting systems \cite{roberts2009impurity,santamore2011multi,blinova2013two,casteels2013bipolarons}. Effective quasiparticle-quasiparticle interactions have been investigated using variational methods \cite{devreese2009frohlich,kashirina2010large}, Dyson's equation \cite{utesov2018effective} and a scattering matrix approach \cite{camacho2018bipolarons,bijlsma2000phonon} to name only a few. Besides electron-phonon coupling, other kinds of indirect interactions play a key role in quantum systems, such as e.g. the Ruderman-Kittel-Kasuya-Yosida interaction \cite{ruderman1954indirect,zhou2010strength}, giving rise to complex magnetic phases such as spin glasses \cite{hewson1997kondo}.

In this paper we analyse the effective interaction between two diatomic molecules mediated by a bosonic bath. Unlike electrons or ground-state atoms, the low-energy degrees of freedom for molecules involve rotations, leading to an exchange of angular momentum between the molecule and the bath. Recently, it has been shown that individual molecules interacting with a bosonic bath form angulon quasiparticles -- rigid rotors dressed by a cloud of excitations carrying angular momentum~\cite{schmidt2015rotation,schmidt2016deformation,lemeshko2017molecular,bighin2018diagrammaticprl}. The results of this theory are in good agreement with a wide range of experimental data including static and dynamic molecular properties \cite{lemeshko2017quasiparticle, shepperson2017laser, cherepanov2017fingerprints, shepperson2017strongly,cherepanov2019far}. In addition to this, it was shown that due to the non-Abelian $SO(3)$ algebra and the discrete energy spectrum inherent to rotations, novel phenomena such as effective magnetic monopoles \cite{yakaboylu2017emergence} and anomalous electrostatic screening \cite{yakaboylu2017anomalous} can emerge. During recent years, molecular complexes in He nanodroplets have been created (see e.g. Refs.~\onlinecite{Toennies2004,PhysRevLett.120.113202,doi:10.1021/acs.jpca.9b07302,doi:10.1021/acs.jpca.6b06227}), and techniques to control molecular alignment in helium have been developed \cite{shepperson2017laser, shepperson2017strongly,cherepanov2019far}. These and other experimental advances pave the way to control and enhance chemical reactivity inside superfluids at the microscopic level.

This motivates us to investigate the effective phonon-mediated interactions between two molecules immersed in a bosonic bath. To investigate the system in various parameter regimes, we apply different theoretical approaches based on angulon theory and several approximations, such as a product-state ansatz, a one-phonon-excitation variational approach and a diagonalization scheme based on single angulon basis states.

All approaches we use in this paper suggest the appearance of a correlated state that we call the \emph{biangulon}. It consists of two diatomic molecules that align with respect to each other due to the effective phonon-mediated interaction. We characterize this effective interaction within the Born-Oppenheimer approximation and show that it depends on both the angular momentum quantum number $L$ and the magnetic quantum number $M$ of each of the two molecules and that if favours states whose phonon clouds overlap strongly with the molecules. Within the Pekar approximation \cite{pekar1946local}, we show that two diatomic molecules show a strong alignment in the strong-coupling regime. Subsequently, employing a one-phonon ansatz, we find that the biangulon shows two spectral instabilities in the weak-coupling regime as well as a shift of the angulon instabilities. These features are proposed as experimental signature for the formation of a biangulon. Finally, a diagonalization scheme based on single angulon and bare rotor basis functions is used, to investigate a system, where the coupling between the bath and one of the two impurities is weaker than the one of the other. In this situation we study the transition from separated angulons to a biangulon by calculating the wavefunction and the rotational correlations between the two molecules. 
\section{\label{sec:level2}The model}
\begin{figure}[t]
\centering
\includegraphics[width=0.3\textwidth]{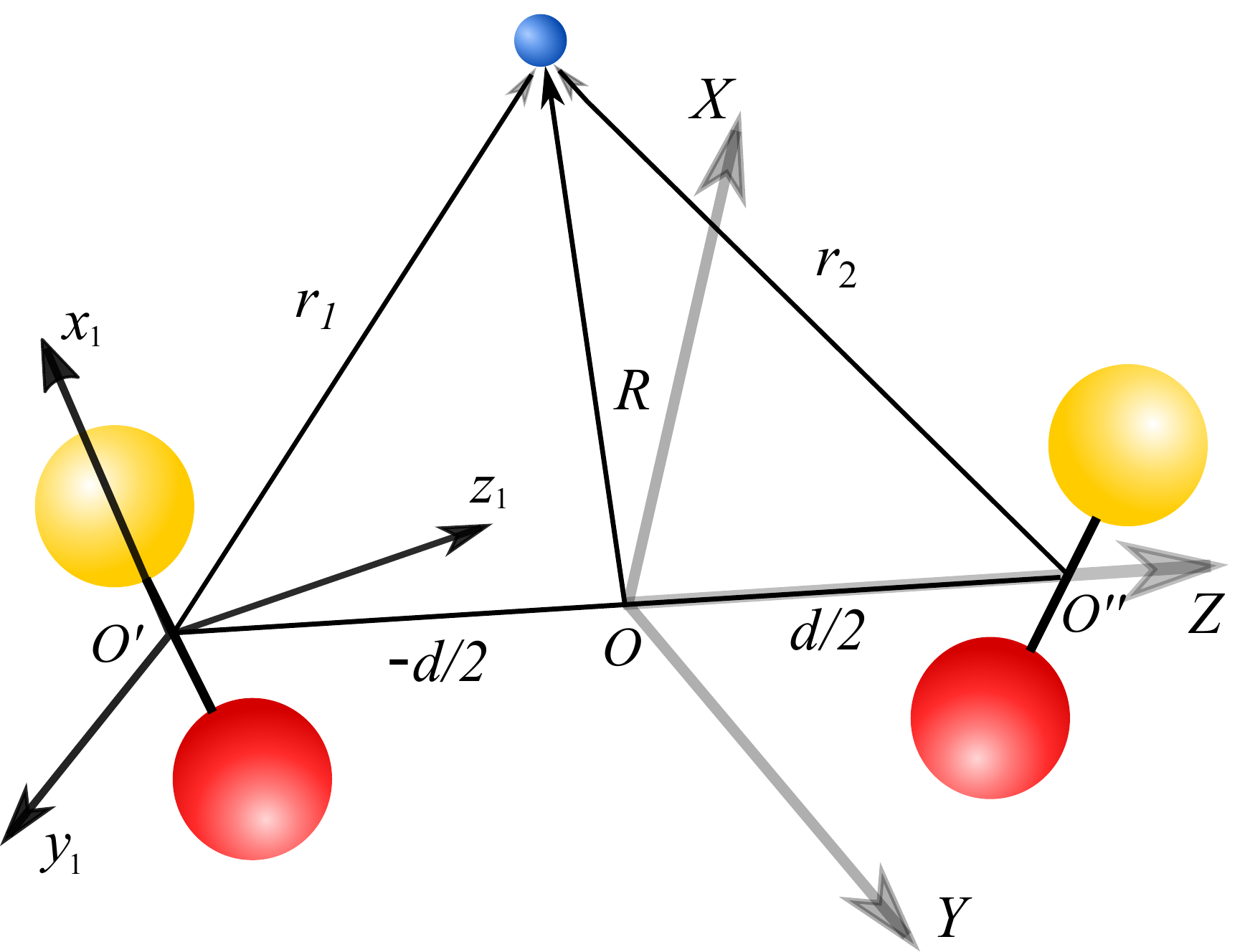}
\caption{Schematic illustration of two  rotating molecular impurities   interacting with a bosonic atom. The origin of the laboratory frame, $\{X, Y, Z\}$, is chosen in the middle between the two molecules on the $Z$-axis. Anisotropic molecule-boson interactions are defined in the molecular coordinate frames labeled by $\{x_i, y_i, z_i\}$ ($i=1, 2$).}
\label{fig:1}
\end{figure}

We consider two rigid linear molecules ($i=1,2$), whose position is fixed in space at $(0, 0, \pm d/2)$ in the laboratory frame with coordinates $\{X, Y, Z\}$, see Fig.~\ref{fig:1}. The rotational kinetic energy of the $i$-th molecule is given by \cite{lemeshko2017molecular}
\begin{equation}
\hat{H}^{(i)}_{\text{mol}} = B_i\hat{\mathbf{J}}^2_i,
\label{eq:hmol}
\end{equation}
where we denote the rotational constant and the angular momentum operator of the $i$-th molecule by $B_i$ and $\hat{\mathbf{J}}_i$, respectively. Here and in the rest of this paper, we assume that the two molecules have the same rotational constant $B=B_1=B_2$. 

The molecules are immersed in a bath of phonons, whose kinetic energy is given by
\begin{equation}
\hat{H}_{\text{bos}} = \sum_{\mathbf{k}}\omega(k)\hat{b}^\dag_{\mathbf{k}}\hat{b}_{\mathbf{k}}.
\label{eq:hbos}
\end{equation}
By $\omega(k)$ with $k = |k|$ we denote the phonon dispersion relation, which will be specified later, and $\hat{b}^\dag_{\mathbf{k}}$, $\hat{b}_{\mathbf{k}}$ with $[\hat{b}_{\mathbf{k}},\hat{b}^{\dagger}_{\mathbf{q}}] =(2 \pi)^3 \delta(\mathbf{k}-\mathbf{q})$ are the usual bosonic creation and annihilation operators of an excitation with momentum $\mathbf{k}$, respectively. 

We assume the coupling between the impurities and the phonons to be linear in the phonon field. In the molecular coordinate frame with coordinates $\{x_i,y_i,z_i\}$, see Fig.~\ref{fig:1}, their interaction is therefore given by
\begin{equation}
\hat{H}^{(i)}_{\text{int}} = \sum_{\mathbf{k}} V(\mathbf{k},\hat{\theta}_i,\hat{\phi}_i)\hat{b}^\dag_{\mathbf{k}} + \text{H.c.},
\label{eq:hint}
\end{equation}
with the effective interaction potential $V(\mathbf{k},\hat{\theta}_i,\hat{\phi}_i)$. A detailed microscopic derivation of an effective interaction of the form~\eqref{eq:hint} for the case of an impurity immersed in a Bose-Einstein condensate is presented in Refs.  \onlinecite{schmidt2015rotation,lemeshko2017molecular}. The interaction \eqref{eq:hint} also serves as a reliable phenomenological model for molecules immersed in helium nanodroplets\cite{lemeshko2017quasiparticle, shepperson2017laser, cherepanov2017fingerprints, shepperson2017strongly,cherepanov2019far}. In this paper we focus on intermolecular forces mediated by phonons, and therefore neglect direct molecule-molecule interactions, such as electrostatic, induction, and dispersion potentials~\cite{StoneBook13}, which can, however, be added to the theory in a straightforward manner.  

As schematically depicted in Fig.~\ref{fig:1}, the two molecules are placed along the $Z$ axis at the points $(0,0,\pm d/2)$, so that the Hamiltonian of the full system in the laboratory frame, $\{X,Y,Z\}$, is given by
\begin{eqnarray}
& &\hat{H} = B\hat{\mathbf{J}}^2_1 + B\hat{\mathbf{J}}^2_2 + \sum_{\mathbf{k}}\omega(k)\hat{b}^\dag_{\mathbf{k}}\hat{b}_{\mathbf{k}}
\label{eq:hbiang2} \\
& &+ \sum_{\mathbf{k}}\left[V(\mathbf{k},\hat{\theta}_1,\hat{\phi}_1)e^{-i\frac{\mathbf{k}\cdot\mathbf{d}}{2}}+ V(\mathbf{k},\hat{\theta}_2,\hat{\phi}_2)e^{i\frac{\mathbf{k}\cdot\mathbf{d}}{2}}\right]\hat{b}^\dag_{\mathbf{k}} +  \text{H.c.}.
\nonumber
\end{eqnarray} 
To obtain this representation, we applied the translation operator $\hat{T}(\mathbf{r}) = \exp(-i{\mathbf{r}}\cdot\sum_{\mathbf{k}}\mathbf{k}\hat{b}_{\mathbf{k}}^\dag\hat{b}_{\mathbf{k}})$ to the interaction term in Eq. \eqref{eq:hint}, see also Ref. \onlinecite{yakaboylu2018theory}. 
\section{\label{sec:ang}Angulons and biangulons}
If the distance between the two molecules is sufficiently large, each single impurity can be described by a (appropriately translated) Hamiltonian of the form
 \begin{equation}
     \hat{H}^{(i)}=B\hat{\mathbf{J}}^2_i + \sum_{\mathbf{k}}\omega(k)\hat{b}^\dag_\mathbf{k}\hat{b}_\mathbf{k} + \sum_{\mathbf{k}}V(\mathbf{k},\hat{\theta}_i,\hat{\phi}_i)\hat{b}^\dag_\mathbf{k}+\text{H.c.}
 \end{equation}
describing one rotating impurity immersed in the bosonic bath. It has been shown that the above Hamiltonian allows for a description of the rotating impurity in terms of the angulon quasiparticle in many different experimental settings, ranging from ultracold gases \cite{Midya2016} to helium nanodroplets \cite{lemeshko2017quasiparticle}. The concept of the biangulon quasiparticle we propose in this paper is based on the analysis of the Hamiltonian \eqref{eq:hbiang2}. If the two molecules come close enough together they will be subject (as we will see below) to an effective attractive interaction mediated by the bosonic bath. As a consequence, a correlated state, where both rotors are dressed by the bath and at the same time strongly interact with each other, is formed. This correlated state is characterized by the fact that the two rotating molecules align with respect to each other such that the phonon cloud of each molecule overlaps with the other molecule. This behavior is very different from that of two uncorrelated (or weakly correlated) angulons and can be found in the regimes of moderate and strong coupling. 

The system of the two impurities placed at $(0,0,\pm d/2)$  is rotationally symmetric around the $z$ axis, and hence the biangulon quasiparticle can be characterized by the magnetic quantum number $M$ of the entire system. This should be compared to the angulon, where one has a full rotational symmetry and the total angular momentum $L$ is also a good quantum number.

In the case of two polarons a bipolaron can form if the effective interaction between the two impurities allows for a bound state \cite{salje2005polarons}. Since our molecules have a frozen center-of-mass motion, this definition is clearly not appropriate, and we therefore opt for the definition above. In practice we expect the two definitions to coincide if the effective attractive interaction between the molecules allows for a bound state.

In the following Sections we will quantitatively study the above two-impurity system and its properties with various theoretical approaches and in different parameter regimes.

\section{\label{sec:level3}Product-state ansatz}
\subsection{\label{subsec:IIIa} Phonon-mediated intermolecular forces}

When the characteristic timescale of the phonons is much shorter than that of molecular rotations, one can assume that the phonons adjust instantaneously to changes of the molecular orientation in space and a Born-Oppenheimer approximation is valid. This corresponds to a product state ansatz
\begin{equation}
    |\psi_\text{b}\rangle = |\text{mol} \rangle \hat{U}|0\rangle.
    \label{eq:pekar}
\end{equation}
Analogous to the Pekar ansatz for polarons~\cite{pekar1946local,Devreese15} the unitary $\hat{U}$ in the above equation is chosen as
\begin{equation}
\label{eq:U}
    \hat{U}= \exp\left[-\sum_{\mathbf{k}}\left(\frac{\langle\hat{f}\rangle}{\omega(k)}\hat{b}^\dag_{\mathbf{k}}-\frac{\langle\hat{f}\rangle^*}{\omega(k)}\hat{b}_{\mathbf{k}}\right)\right],
\end{equation}
where
\begin{equation}
    \langle\hat{f}\rangle = \langle\text{mol}|V(\mathbf{k},\hat{\theta}_1,\hat{\phi}_1)e^{-i\frac{\mathbf{k}\cdot\mathbf{d}}{2}}+ V(\mathbf{k},\hat{\theta}_2,\hat{\phi}_2)e^{i\frac{\mathbf{k}\cdot\mathbf{d}}{2}}|\text{mol}\rangle.
\end{equation}
We stress that the description of the bath in terms of the coherent state $\hat{U} | 0 \rangle$ in Eq.~\eqref{eq:pekar} takes an arbitrary number of phonon excitations into account.

Since we are interested in angular momentum exchange between the molecules and the environment, it is convenient to expand the bosonic field operators in the angular momentum basis as
\begin{equation}
\hat{b}^\dag_{\mathbf{k}} = \frac{(2 \pi)^{3/2}}{k} \sum_{\lambda \mu} \hat{b}^\dag_{k \lambda \mu} i^{\lambda} Y_{\lambda \mu}^*(\theta_k,\phi_k)
\end{equation}
see e.g. \onlinecite{lemeshko2017molecular}. Here $\hat{b}^\dag_{k \lambda \mu}$ creates a phonon with radial momentum $k$, angular momentum $\lambda$ and projection onto the z-axis $\mu$. By $Y_{\lambda\mu}(\theta_k,\phi_k)$ we denote the spherical harmonics. Additionally, $\theta_k$, $\phi_k$ are the angles determined by $\mathbf{k}$ in spherical coordinates and $k$ denotes its absolute value. The inverse relation reads
\begin{equation}
\hat{b}^\dag_{k \lambda \mu} = \frac{k}{(2 \pi)^{3/2}} \int \mathrm{d}  \phi_{\mathrm{k}} \mathrm{d} \theta_{\mathrm{k}} \sin(\theta_{\mathrm{k}}) \hat{b}^\dag_{\mathbf{k}} i^{-\lambda} Y_{\lambda \mu}(\theta_{\mathrm{k}}, \phi_{\mathrm{k}}).
\end{equation}
We also write the interaction potential as
\begin{equation}
V(\mathbf{k},\hat{\theta}_i,\hat{\phi}_i) = \sum_{\lambda\mu}(2\pi)^{3/2}i^{-\lambda} \frac{U_{\lambda}(k)}{k}Y_{\lambda\mu}(\theta_k,\phi_k) Y^*_{\lambda\mu}(\hat{\theta}_i,\hat{\phi}_i),
\label{eq:mol-bos}
\end{equation}
where the potential has been expanded in partial wave components $U_{\lambda}(k)$\cite{lemeshko2017molecular}.

For specific molecular rotational states $|\text{mol}\rangle = |L_1M_1L_2M_2\rangle$, where $L_i$ and $M_i$ denote the angular momentum quantum number and the magnetic quantum number of the $i$-th molecule, the energies $E_{\text{BA}}=\langle\psi_{\text{b}}|\hat{H}|\psi_{\text{b}}\rangle$ of the Hamiltonian \eqref{eq:hbiang2} can be readily calculated. Applying the same approach to a single molecular impurity in a state state $|L_iM_i\rangle$, one obtains the energy $E_{\text{A}}^{(i)}$ of one angulon quasiparticle. In order to measure the strength of the interaction between two angulons we define the \emph{effective angulon-angulon interaction} as
\begin{equation}
    \Delta E = E_{\text{BA}} - E_{\text{A}}^{(1)} - E_{\text{A}}^{(2)}.
    \label{eq:VAA}
\end{equation}
A similar definition for two polarons can be found in Refs.  \onlinecite{devreese2009frohlich,kashirina2010large}.

Here and in what follows we choose parameters that are well suited to describe two molecular impurities immersed in a bath of superfluid $^4\text{He}$. More precisely, we choose the phonon dispersion relation as $\omega(k)=\sqrt{\epsilon(k)(\epsilon(k)+2g_{\text{bb}}n)}$, where $\epsilon(k)=k^2/2m$, $g_{bb} = 4 \pi a /m$ with the scattering length $a$ and the mass $m$ of the Helium atoms. The function $\omega(k)$ is an approximation to the dispersion relation of sound waves in liquid helium that is valid at low momenta. By $n$ we denote the density of the Helium atoms. To describe a typical atom-molecule interaction, we choose 
\begin{equation}
	U_{\lambda}(k)=u_\lambda\left(\frac{8nk^2\epsilon(k)}{[\omega(k)(2\lambda+1)]} \right)^{1/2}\int drr^2f_\lambda(r)j_\lambda(kr)
	\label{eq:6}
\end{equation}
with Gaussian form factors $f_\lambda(r)=(2\pi)^{-3/2}e^{-r^2/(2r^{2}_{\lambda})}$. Here $j_\lambda(kr)$ denotes the spherical Bessel function. The coupling strengths and the potential radii are chosen as $u_0=u_2=218 B$, $u_{\lambda}=0$ if $\lambda \neq 0,2$ and $r_0=r_2= 1.5(mB)^{-1/2}$, respectively \cite{stone2013theory,schmidt2015rotation}. We also choose $a=3.3 (m B)^{-1/2}$, which reproduces the speed of sound in superfluid Helium for a molecule whose rotational constant is $B = 2\pi\times 1$ GHz\cite{donnelly1998observed,schmidt2015rotation}.

\begin{figure}[t]
	\centering
	\includegraphics[width=0.5\textwidth]{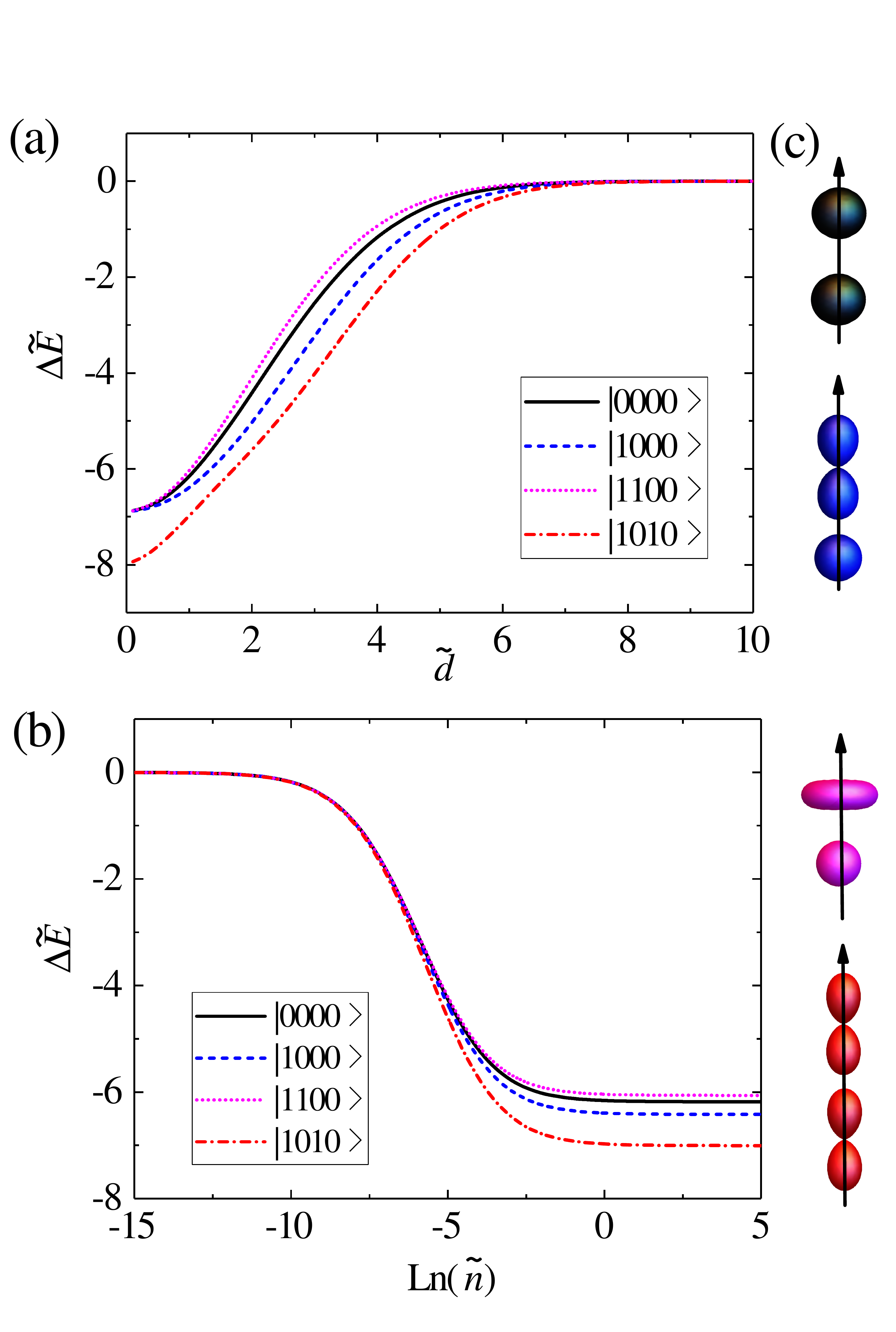}
	\caption{Dimensionless angulon-angulon interaction $\Delta\tilde{E} = \Delta E/B$, Eq.~\eqref{eq:VAA}, calculated using the product state ansatz, Eq.~\eqref{eq:pekar}, as a function of (a) the dimensionless molecule-molecule distance  $\tilde{d}=d(mB)^{-1/2}$ and (b) the dimensionless bath density $\tilde{n}=n(mB)^{-3/2}$. We have chosen $\tilde{n} = 1$ for the bath density in (a) and $\tilde{d}=1$ for the distance between the molecules in (b). The black solid line, blue dashed line, magenta dots, and red dashed dots have been computed with the molecular states $|L_1M_1L_2M_2\rangle = |0000\rangle$, $|1000\rangle$, $|1100\rangle$ and $\vert 1010\rangle$, respectively. The squared absolute value of the wave functions related to the different molecular states (with colors as introduced in the legend) are schematically shown in (c). For more information see the text.}
	\label{fig:pekar_vaa}
\end{figure} 
In Fig.~\ref{fig:pekar_vaa} we show the dimensionless effective interaction $\Delta\tilde{E}=\Delta E/B$ as a function of (a) the dimensionless molecule-molecule distance $\tilde{d}=d(mB)^{-1/2}$ and (b) the dimensionless bath density $\tilde{n}=n(mB)^{-3/2}$. The squared absolute value of the wave functions related to the different molecular states (with colors as introduced in the legend) are schematically shown in (c). In subgraph (a) the density is $\tilde{n}=1$ and in (b) the molecule-molecule distance is fixed as $\tilde{d}=1$. When the two molecules are placed far away from each other or when the surrounding bath is sufficiently dilute, the effective interaction is small and the system resembles two separate angulons. 

Outside this parameter regime we observe an attractive interaction between the two rotors ($\Delta\tilde{E}<0$), which results from the linear coupling in the Hamiltonian \eqref{eq:hbiang2}. It is sensitive to the rotational state of the two molecules and takes its largest values when the overlap of the phonon density of each of the two molecules with the other molecule is maximal. Accordingly, it depends also on the magnetic quantum numbers $M_1$ and $M_2$. For example, the effective interaction between molecules in the state $|L_1M_1L_2M_2\rangle=|1000\rangle$ (blue dashed line in Fig. \ref{fig:pekar_vaa}) is stronger than the one between molecules in the state $|1100\rangle$ (magenta dots). The interaction energy of the latter state is even weaker than the one of the state $| 0000 \rangle$ (black solid line) and the state $| 1010 \rangle$ shows the largest interaction energy among the ones that have been considered. See also Fig.~\ref{fig:pekar_vaa}(c) for the shapes of the orbitals related to these molecular states. The anisotropy of the molecular wave function of one molecule is responsible for a similar anisotropy of its phonon cloud. The interaction energy is large if this anisotropy causes a strong overlap of the molecules phonon cloud with the other molecule. In general, the states with $M_1 = M_2 = 0$ show the largest effective interaction. Such an effective interaction clearly favors a biangulon-like behavior if the impurities are sufficiently close. 

The saturation of the effective interaction for large densities $\tilde{n}$ in Fig.~\ref{fig:pekar_vaa} (b) is a consequence of the fact that the phonon dispersion relation $\omega(k)$ and $| \langle\hat{f}\rangle |^2 $ are both proportional to $\sqrt{\tilde{n}}$ in this regime, see Eqs.~\eqref{eq:mol-bos}, \eqref{eq:6} and Eq.~\eqref{eq:Pekarfunct} in Section~\ref{subsec:IIIb} below. The states $| 1,1,1,1 \rangle$ and $| 1,1,1,-1 \rangle$ have the same interaction energy. That is, the effective interaction  is not sensitive to whether the two molecules rotate in the same or in opposite directions. Since both molecules have the same rotational constant $B$, one obtains the same result if their quantum numbers are exchanged.

\subsection{\label{subsec:IIIb} Relative molecular orientation in the ground state}

In this Section we study the ground state of two molecules immersed in the bath of phonons within the Pekar approximation. Accordingly, we minimize the expectation of the Hamiltonian \eqref{eq:hbiang2} over the molecular part of the wave function in \eqref{eq:pekar}, similar to Ref. \onlinecite{pekar1946local}. This approximation is expected to be valid in the strong-coupling regime \cite{Pekarderivation1,Pekarderivation2}.

More precisely, we expand the molecular wave function in angular momentum eigenfunctions as
\begin{equation}
  |\text{mol}\rangle= \sum_{L_1,M_1,L_2,M_2} s_{L_1,M_1,L_2,M_2}|L_1M_1\rangle|L_2M_2\rangle.
    \label{eq:pekar_impurity}
\end{equation}
In the following, we abbreviate $c = (L_1,M_1,L_2,M_2)$. When we insert \eqref{eq:pekar_impurity} into \eqref{eq:pekar} and compute with this wave function the expectation value of $\hat{H}$, Eq.~\eqref{eq:hbiang2}, we obtain the Pekar functional 
\begin{equation}
    \mathcal{E}_{\mathrm{BA}}(s) = \sum_{c}\bigg( B[L_1(L_1+1) + L_2(L_2+1)]|s_{c}|^2 - 
    \sum_{\mathbf{k}}\frac{|\langle\hat{f}\rangle|^2}{\omega(k)}\bigg), \label{eq:Pekarfunct} 
\end{equation}
as well as the biangulon energy
\begin{equation}
	E_{\mathrm{BA}} = \min_{\sum_c | s_c |^2 = 1} \mathcal{E}_{\mathrm{BA}}(s).
	\label{eq:pekar_EBA}
\end{equation}
Similarly, we find the energy $E_{\mathrm{A}}$ of one impurity within the Pekar approximation and we have 
\begin{equation}\label{eq:5}
	\Delta E = E_{\mathrm{BA}} - 2 E_{\mathrm{A}}.
\end{equation} 

To minimize $\mathcal{E}_{\mathrm{BA}}(s)$ numerically, we introduce the cut-off $L_1, L_2, |M_1|, |M_2| \leq 4$ for the values of the angular momentum quantum number. The minimization is then carried out with a stochastic simulated annealing procedure based on moves that can reach any allowed value of the variational coefficients\cite{das2005quantum}. More details on the procedure can be found in Appendix \ref{sec:appendix}.

For a better understanding of the resulting state we also consider the alignment cosine 
\begin{align}
\langle\cos^2\theta_1 \rangle =\sum_{c,c'} \tilde{s}^*_{c'} \tilde{s}_{c} \langle L_1'M_1'|\cos^2\theta_1|L_1 M_1 \rangle \delta_{L_2',L_2}\delta_{M_2'M_2},
\label{eq:alignment}
\end{align}
where $\tilde{s}$ denotes the minimizer of $\mathcal{E}_{\mathrm{BA}}$. The expectation value on the left-hand side is taken with respect to the state $|\psi_\text{b}\rangle$ in Eq.~\eqref{eq:pekar}, where the molecular wave function is replaced by the wave function in Eq.~\eqref{eq:pekar_impurity} with coefficients given by $\tilde{s}$. From our computations we see that the minimizer of $\mathcal{E}_{\mathrm{BA}}$ is a product state that is symmetric in the two impurities (for the case $B=B_1=B_2$). This implies $\langle\cos^2\theta_1 \rangle = \langle\cos^2\theta_2 \rangle$, and hence we can use Eq.~\eqref{eq:alignment} to measure the anisotropy of the molecular orientation of both molecules.

\begin{figure}[t]
	\centering
	\includegraphics[width=0.45\textwidth]{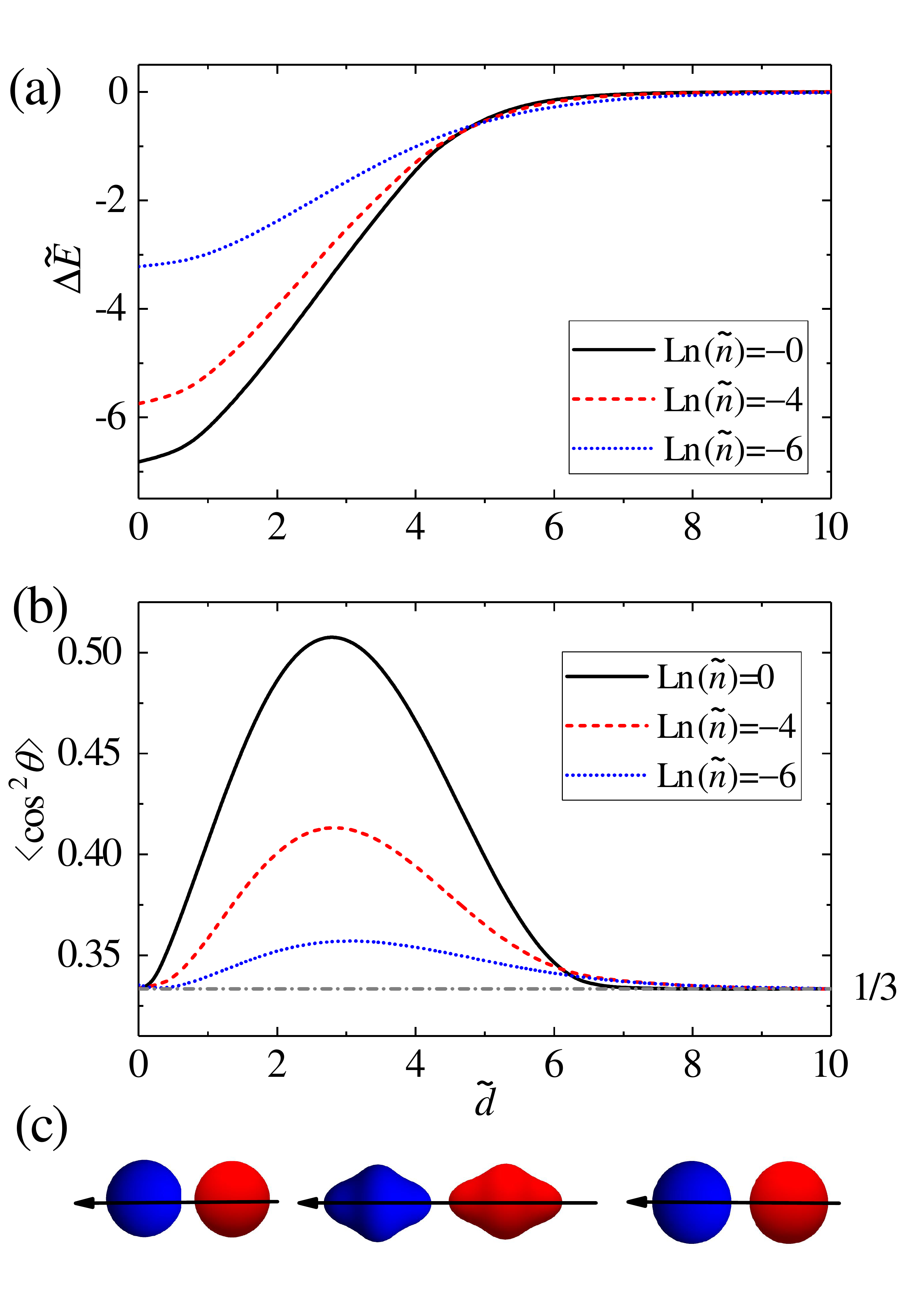}
	\caption{(a) Dimensionless effective interaction $\Delta \tilde{E}=\Delta E_{\mathrm{BA}} /B$, Eq.~\eqref{eq:5}, and (b) the alignment cosine $\langle\cos^2\theta_1 \rangle$, Eq.~\eqref{eq:alignment}, of one of the molecules computed within the Pekar approximation as a function of the dimensionless molecule-molecule distance $\tilde{d}=d(mB)^{-1/2}$ for different dimensionless bath densities $\tilde{n}=n(mB)^{-3/2}$. In (c) we show schematic figures of the wave functions of the two molecules for the parameters $\tilde{n} = 1$ and $\tilde{d} = 0.3$ (left picture), $\tilde{d} = 3$ (picture in the middle) and $\tilde{d} = 8$ (right picture). For more details see the text.}
	\label{fig:min}
\end{figure}
In Fig.~\ref{fig:min} we show (a) the dimensionless effective interaction $\Delta \tilde{E}=\Delta E/B$ and (b) the alignment cosine $\langle\cos^2\theta_1 \rangle$ as a function of the dimensionless molecule-molecule distance $\tilde{d}=d(mB)^{-1/2}$. The dimensionless bath density $\tilde{n}=n(mB)^{-3/2}$ is chosen as $\ln(\tilde{n}) = 0$ (black solid line), $\ln(\tilde{n}) = -4$ (red dashed line), and $\ln(\tilde{n}) = -6$ (blue dot line). In (c) we show schematic figures of the wave functions of the two molecules for the parameters $\tilde{n} = 1$ and $\tilde{d} = 0.3$ (left picture), $\tilde{d} = 3$ (picture in the middle) and $\tilde{d} = 8$ (right picture). As one would expect, the effective interaction is an increasing function of the bath density and a decreasing function of the distance between the impurities. 

For large distances the ground state is given by the impurity wave function $\vert L_1M_1L_2M_2\rangle=|0000\rangle$ and the two molecules form two isolated angulon quasiparticles with no preferential orientation. In this case the alignment cosine equals $1/3$, see Fig.~\ref{fig:min}(b). If they come closer together, contributions with nonzero angular momentum and $M_i = 0$ for $i=1,2$ become relevant, compare with (c). This is in accordance with the analysis in Section~\ref{subsec:IIIa}, see Fig.~\ref{fig:pekar_vaa}, where we found that such states maximize the overlap of the phonon cloud of each of the two impurities with the other impurity, and therewith also their attractive interaction. This behavior is also captured by the alignment cosine, which takes its largest values around $\tilde{d} = 3$. In this region the two impurities form a biangulon quasiparticle, which is characterized by the fact that their relative orientation is strongly correlated and that their phonon densities are highly anisotropic. 

If the distance is further decreased the phonon clouds are already substantially overlapping with the molecules if the molecular wave function is almost rotationally symmetric and an anisotropy of the molecular orientation is no longer beneficial. This is indicated by $\langle\cos^2\theta_1 \rangle \to 1/3$ for small $\tilde{d}$. In other words, the short distance behavior of the two impurities is a perturbation of the extreme case $\tilde{d}=0$, where the model has full rotational symmetry. In practice one would need to introduce a repulsive interaction between the two molecules in order to describe the relevant physics in the regime of very small $\tilde{d}$ correctly. This, however, goes beyond the scope of the present paper.

Finally, let us note that the Gaussian form factors and our choice of the dispersion relation imply that the effective interaction is an exponentially decaying function of the distance $\tilde{d}$. This can be seen as follows: We have already noted that the molecular wave function is given by $s_{c}=\delta_{L_1,0}\delta_{M_1,0}\delta_{L2,0}\delta_{M_2,0}$ if $\tilde{d}$ is chosen sufficiently large, compare with Fig.~\ref{fig:min}(b). In this case we can write the effective interaction as
	\begin{align}
	&\Delta E = \label{eq:VAAanalytic} \\ &-\frac{1}{2\pi}\sum_{k}\frac{U_0^2(k)}{\omega(k)}\left[\sum_\lambda(2\lambda+1)j_\lambda(kd/2 )^2(1+(-1)^\lambda)-1\right]. \nonumber	
	\end{align}
With $\sum_\lambda(2\lambda+1)j_\lambda(x)^2=1$ and $\sum_\lambda(-1)^\lambda(2\lambda+1)j_\lambda(x)^2=\frac{\sin{2x}}{2x}$~\cite{abramowitz1965handbook}, Eq.~\eqref{eq:VAAanalytic} simplifies to
	\begin{equation}
	\Delta E  = -\frac{1}{2\pi}\sum_k\frac{U^2_0(k)}{\omega(k)}\frac{\sin(kd)}{kd}.
	\label{eq:VAAanaly2}
	\end{equation}
Our choice of the form factor in Sec.~\ref{subsec:IIIa} implies
\begin{equation}
	U_0(k) = u_0 \left( \frac{8 n k^2 \epsilon(k)}{\omega(k)} \right)^{1/2} \frac{r_0^4 e^{-r_0^2 k^2/2}}{2^{5/2} \pi}. 
	\label{eq:1}
\end{equation}
We insert \eqref{eq:1} and $\omega(k)$ from Sec.~\ref{subsec:IIIa} into \eqref{eq:VAAanaly2}. Integration by parts and an application of Ref.~\onlinecite{ReedSimon1980}, Theorem~IX.13 therein, then shows the claim.  

It should also be noted that for the Fr\"ohlich parameters $\omega(k)=\omega_0$ and  $U_0(k)=U_0$ one finds the well-known behavior \cite{Pekarderivation2}
	\begin{equation}
	\Delta E\propto\frac{1}{\tilde{d}}.
	\end{equation}

\section{\label{sec:level4}One-phonon-excitation variational ansatz}
The product-state-ansatz of Section \ref{sec:level3} describes molecular impurities dressed by an arbitrary number of phonons in a coherent state (cf. Eq.~\eqref{eq:U}). Minimization over the impurity wave function yields the Pekar approximation, which is expected to be valid for strong molecule-bath interactions \cite{Pekarderivation1,Pekarderivation2}. When the molecule-bath interaction is weak, however, we expect only a small number of phonons to be excited. It is the aim of the present Section to investigate such a situation in detail.

More precisely, we are going to use a one-phonon-excitation variational ansatz, that is, we will allow for at most one phonon in the system. Such an ansatz has been successfully applied in several different contexts, see Refs.  \onlinecite{chevy2006universal,lan2014single,schmidt2015rotation}. For a system of two rotating molecules immersed in a bosonic bath this variational ansatz reads
\begin{eqnarray}
    |\psi_\text{1-ph}\rangle = g|L_1M_1\rangle|L_2M_2\rangle|0\rangle +  \sum_{c}\beta_{c}|j_1m_1\rangle|j_2m_2\rangle\hat{b}^\dag_{\mathbf{k}}|0\rangle,
    \nonumber\\
    \label{eq:chevy}
\end{eqnarray}
where $c = (j_1,m_1,j_2,m_2,\mathbf{k} )$, and the sum over $\mathbf{k}$ is actually an integral. The variational coefficients $g$ and $\beta_{c}$ are chosen such that the magnetic quantum number $M = M_1 + M_2$ of the whole system is a good quantum number and such that $|g|^2 + \sum_{c}|\beta_{c}|^2 = 1$ holds. The first term in Eq.~(\ref{eq:chevy}) describes two free rotors and a bosonic bath in its vacuum state. In the second term a phonon with momentum $\mathbf{k}$ is excited and introduces correlations between the two molecules and the bath. We expect the ansatz \eqref{eq:chevy} to be a good approximation in situations where the helium density $\tilde{n}$ is sufficiently dilute and/or when the distance between the two impurities is such that we still have moderate correlations between them. Accordingly, it describes either a weakly correlated biangulon or two weakly interacting angulons. 

When we compute the expectation value of $\hat{H}$ \eqref{eq:hbiang2} in the state $|\psi_\text{1-ph}\rangle$ and minimize the functional $F(\psi_\text{1-ph})=\langle \psi_\text{1-ph} |\hat{H}-E| \psi_\text{1-ph} \rangle$ with respect to the variational coefficients, we obtain the self-consistent equation
\begin{equation}
    E_{\text{BA}} = BL_1(L_1+1) + BL_2(L_2+1) - \Sigma^{\text{BA}}_{L_1M_1L_2M_2}(E_{\text{BA}})
    \label{eq:dyson}
\end{equation}
for the energy $E_{\text{BA}}$. Here the self-energy $\Sigma^{\text{BA}}_{L_1M_1L_2M_2}(E_{\text{BA}})$ is given by
\begin{eqnarray}
&&\Sigma^{\text{BA}}_{L_1M_1L_2M_2}(E_{\text{BA}}) =
\nonumber\\
&&\sum_{k\lambda j_1}\frac{2\lambda+1}{4\pi}\frac{U^2_\lambda(k)\left[C^{j_10}_{L_10,\lambda0}\right]^2}{Bj_1(j_1+1)+BL_2(L_2+1)+\omega(k)-E_{\text{BA}}}
\nonumber\\
&&+\sum_{k\lambda j_2}\frac{2\lambda+1}{4\pi}\frac{U^2_\lambda(k)\left[C^{j_20}_{L_20,\lambda0}\right]^2}{BL_1(L_1+1)+Bj_2(j_2+1)+\omega(k)-E_{\text{BA}}}
\nonumber\\
&&+\sum_{k\lambda\lambda'\mu}\frac{C^{L_10}_{L_10,\lambda0}C^{L_1M_1}_{L_1M_1,\lambda\mu}C^{L_20}_{L_20,\lambda'0}C^{L_2M_2}_{L_2M_2,\lambda'\mu}\Gamma_{\lambda,\lambda'}(\mathbf{k},d)}{BL_1(L_1+1)+BL_2(L_2+1)+\omega(k)-E_{\text{BA}}}
\nonumber\\
\label{eq:SigmaBA}
\end{eqnarray}
and
\begin{eqnarray}
   && \Gamma_{\lambda,\lambda'}(\mathbf{k},d) =
   \nonumber\\
   && i^{\lambda-\lambda'}\sqrt{\frac{(2\lambda+1)(2\lambda'+1)}{(4\pi)^2}}U_\lambda(k)U_{\lambda'}(k)\times
   \nonumber\\
   &&\int d\phi_k\int d\theta_k\sin{\theta_k}\left[e^{i\mathbf{k}\cdot\mathbf{d}}Y_{\lambda,\mu}(\theta_k,\phi_k)Y^*_{\lambda',\mu}(\theta_k,\phi_k) + \text{c.c.}\right].
   \nonumber\\
\end{eqnarray}
By $C^{l_1m_1}_{l_2m_2,l_3m_3}$ we denote the Clebsch-Gordan coefficients \cite{varshalovich1988quantum}.

\subsection{\label{subsec:IVa} The spectral function and instabilities}
\begin{figure}[t]
	\centering
	\includegraphics[width=0.5\textwidth]{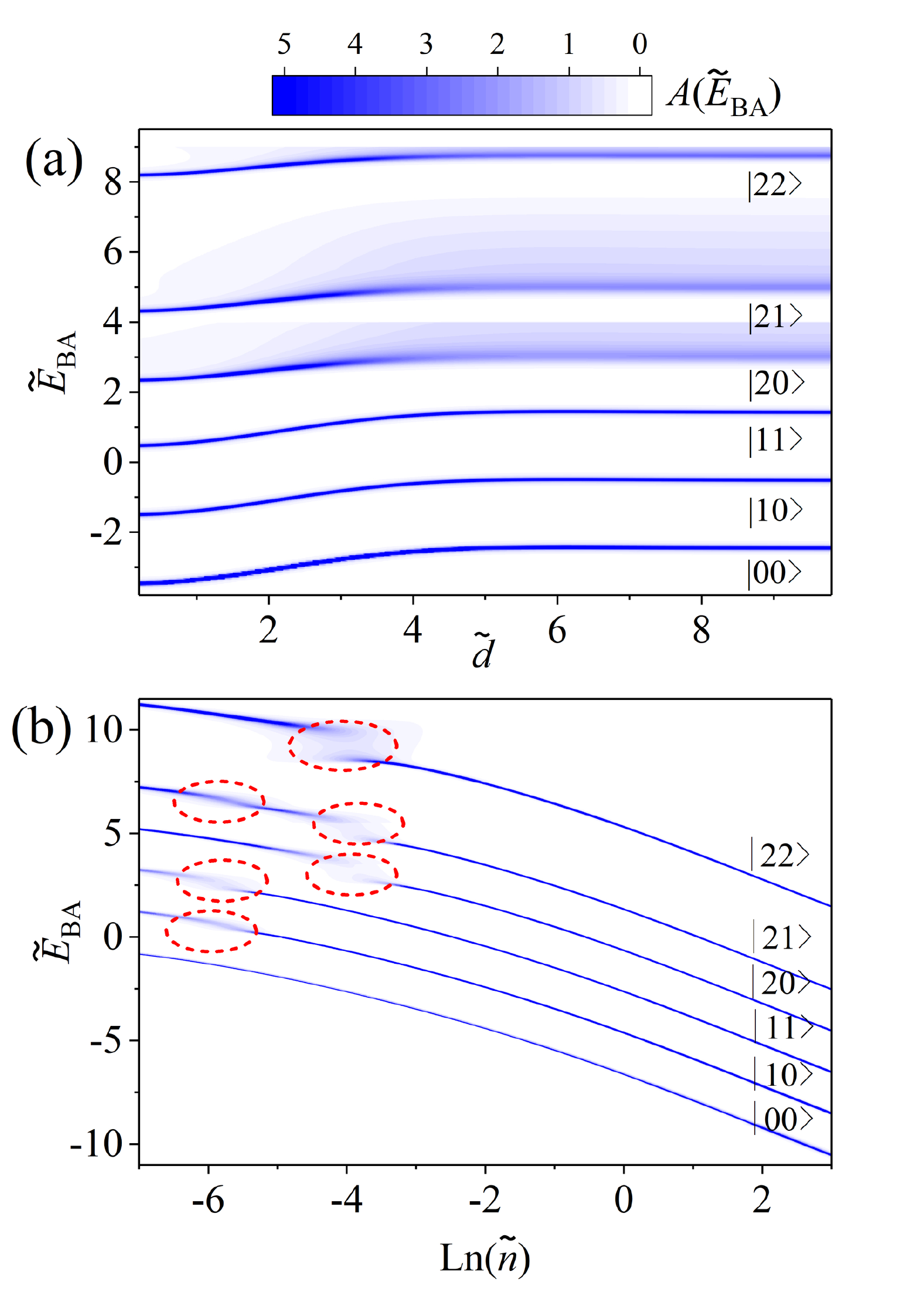}
	\caption{Spectral function $\text{A}_{j_1j_2}(\tilde{E}_{\mathrm{BA}})$, Eq.~\eqref{eq:spec}, of the biangulon  as a function of the dimensionless energy $\tilde{E}_{\mathrm{BA}}$ and (a) the dimensionless molecule-molecule distance $\tilde{d}$ as well as (b) the dimensionless bath density $\tilde{n}$ for different angular momentum states $L_1$ and $L_2$ with  $M_1=0=M_2$. The states are labeled according to the first term in \eqref{eq:chevy} and we use the notation $| L_1 L_2 \rangle = | L_1, M_1=0, L_2, M_2 = 0 \rangle$. In (a) the bath density is chosen as $\ln(\tilde{n}) = -3$ and the distance in (b) is given by $\tilde{d}=0.6$. Biangulon instabilities are highlighted by  red dotted circles. For details see the text.}
	\label{fig:spec}
\end{figure}

As for a single molecule immersed in a bosonic bath \cite{lemeshko2017molecular}, the self-consistent equation (\ref{eq:dyson}) gives us access to the biangulon spectral function
\begin{equation}
    \text{A}_{L_1L_2}(E)=\text{Im}[G^{\text{BA}}_{L_1M_1L_2M_2}(E+i0^+)],
    \label{eq:spec}
\end{equation}
where
\begin{align}
    &G^{\text{BA}}_{L_1M_1L_2M_2}(E) = \label{eq:Green_function} \\
    &\hspace{1cm} \frac{1}{BL_1(L_1+1)+BL_1(L_1+1)-E-\Sigma^{\text{BA}}_{L_1M_1L_2M_2}(E)}, \nonumber
\end{align}
denotes the retarded Green's function, and therewith to the energy spectrum of the system. 

One of the most striking features of the angulon quasiparticle is the onset of an intermediate instability regime, where resonant transfer of angular momentum between the molecule and the bath drastically decreases the quasiparticle weight\cite{schmidt2015rotation}. This phenomenon has been observed experimentally \cite{cherepanov2017fingerprints}. In order to make our results comparable to the case of one molecular impurity, we choose in this Section the same parameters as in Fig.~2 in Ref.~\onlinecite{schmidt2015rotation}. In Fig.~\ref{fig:spec} we study the biangulon spectral function \eqref{eq:spec} as a function of the dimensionless energy $\tilde{E}_{\mathrm{BA}}$ and (a) the dimensionless molecule-molecule distance $\tilde{d}$ as well as (b) the dimensionless bath density $\tilde{n}$. In (a) we have chosen $\ln(\tilde{n})=-3$, while $\tilde{d}=0.6$ in (b). States are labeled according to the first term in \eqref{eq:chevy}. The biangulon instabilities are highlighted by the red dotted circles. The degeneracy of different $M = M_1 + M_2$ states is lifted by the interaction. To keep the figures accessible, we, however, only consider state with $M_1 = 0 = M_2$ here. This is on the one hand because the quasiparticle instabilities for states with $M_1, M_2 \neq 0$ are very similar to the ones for states with $M_i=0$, and on the other hand because their energies are very close. 

In Fig.~\ref{fig:spec}(a) we see that the biangulon instabilities are only slowly changing with the distance $\tilde{d}$ between the two impurities and appear in a wide region of distances. In this regime a description of the system in terms of the biangulon quasiparticle, or for larger distances in terms of two separate angulons, breaks down. For larger distances this can be explained as follows: The two impurities are weakly interacting and therefore almost independent. If the parameters are such that one of the two impurities experiences an angulon instability the quasiparticle picture breaks down and a further increase of the molecule-molecule distance does not change this situation. 

We note that the instability region, as a function of the adimensional density $\tilde{n}$, has approximately the same size as in the single angulon case, see Fig.~2 in Ref.~\onlinecite{schmidt2015rotation}. We observe, however, that the instability for the biangulon appears at lower densities. For instance, the instability of a single angulon in the molecular state $|LM \rangle = |10 \rangle$ is located around $\ln(\tilde{n})=-5$, see Fig.~2 in Ref.~\onlinecite{schmidt2015rotation}, while Fig.~\ref{fig:spec}(b) shows that the instability is shifted to the region around $\ln(\tilde{n})=-6$ when another molecule in the state $|LM \rangle=|0 0\rangle$ is put at a distance $\tilde{d}=0.6$ from the first one. Furthermore, two spectral instabilities can be found in the biangulon spectrum where there is only one in the case of the angulon: In Fig.~\ref{fig:spec}(b) we see a first instability of the state $|L_1L_2\rangle = |21\rangle$ around $\ln(\tilde{n})=-6$ and a second around $\ln(\tilde{n})=-4$. These two instabilities correspond to phonons excited by molecules with different angular momentum quantum number, in this case $L=1$ and $L=2$. We can distinguish the two instabilities because, compared to the situation in Fig.~2 in Ref.~\onlinecite{schmidt2015rotation}, the relevant angulon instabilties are shifted. Both features, the shift of the spectral instabilities and the appearance of a second instability, can be used in experiments as a measure for correlations between the two impurities, and therewith as a signature for the formation of the biangulon quasiparticle.

We note that the spectral instability of the state $|L_1 L_2 \rangle = |10 \rangle$ appears at $\ln(\tilde{n})=-5.2$ if $\tilde{d} = 10$ and not at $\ln(\tilde{n})=-5$, see Fig.~2 in Ref.~\onlinecite{schmidt2015rotation}, as one would expect for two (almost) non-interacting impurities. This shift is a consequence of our one-phonon excitation variational ansatz, which forces the impurities to share one phonon also if they are far apart from each other. The result is a slightly different dressing of the two impurities by the phonon compared to the case of a single angulon (described by a one-phonon variational ansatz) and explains the above deviation. A careful discussion of this effect can be found in the following Section.

\subsection{\label{subsec:IVb}  Effective interaction}
\begin{figure}[t]
	\centering
	\includegraphics[width=0.5\textwidth]{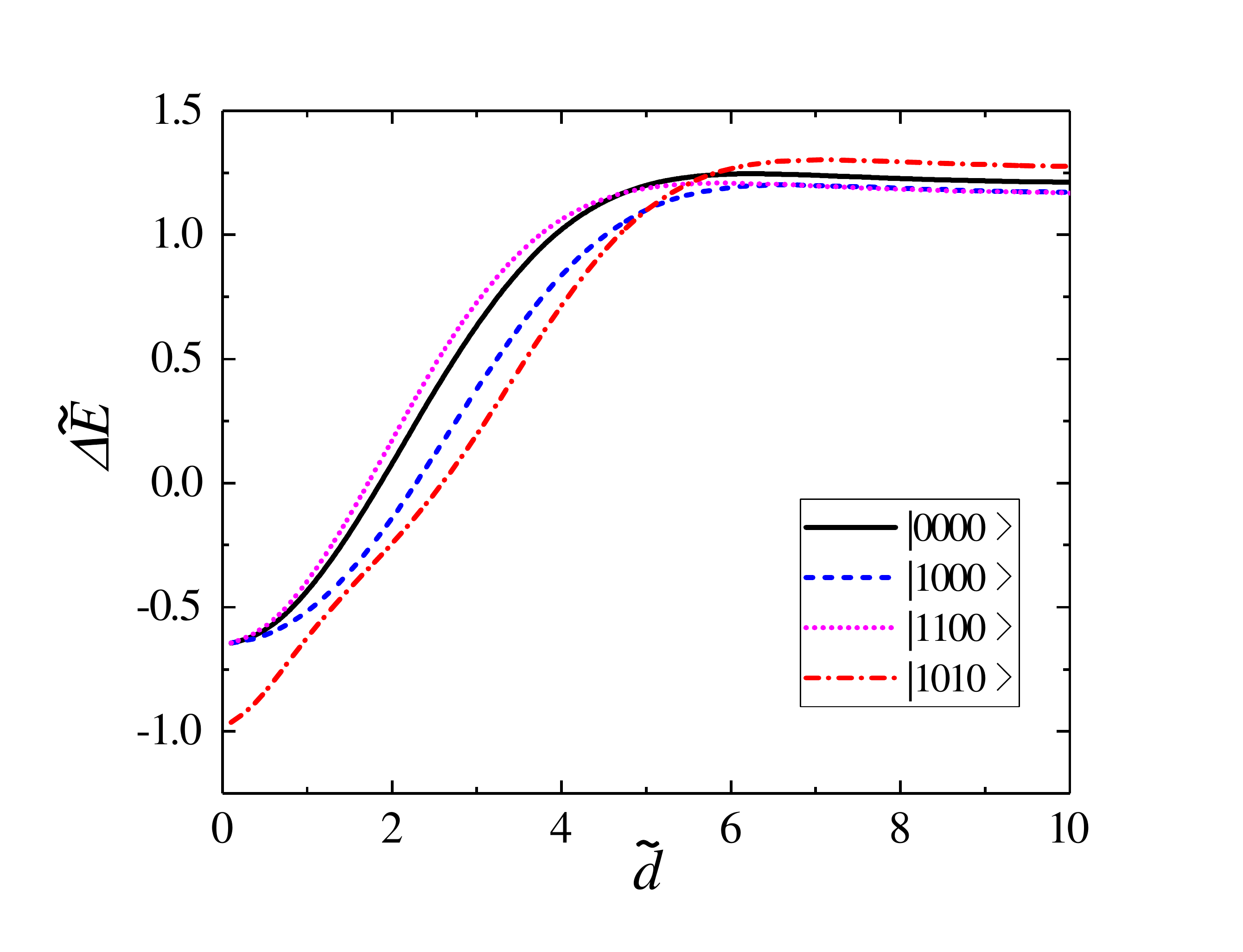}
	\caption{Effective interaction $\Delta \tilde{E}$ obtained with the one-phonon-excitation variational ansatz \eqref{eq:chevy} for molecular states $|L_1M_1L_2M_2\rangle = |1010\rangle$ (red solid line), $|1110\rangle$ (black dot line), and $|1111\rangle$ (blue dashed line) as a function of the dimensionless molecule-molecule distance $\tilde{d}$. States are labeled according to the first term in Eq. \eqref{eq:chevy}. The bath density is chosen such that $\ln(\tilde{n})=0$. For more details see the text.}
	\label{fig:aainter_chevy}
\end{figure}
Let us also consider the effective interaction between the impurities
\begin{equation}\label{eq:effectinteraction1ph}
	\Delta E = E_{\mathrm{BA}} - E_{\mathrm{A}}^{(1)} - E_{\mathrm{A}}^{(2)},
\end{equation}
where $E_{\mathrm{A}}^{(i)}$ denotes the energy of the $i$-th impurity computed with a one-phonon-excitation variational ansatz, see Refs. \onlinecite{schmidt2015rotation,lemeshko2017molecular}. In Fig.~\ref{fig:aainter_chevy} we show $\Delta E$ as a function of the dimensionless distance $\tilde{d}$ for the same quantum numbers as in Fig~\ref{fig:pekar_vaa}, where the Born-Oppenheimer approximation has been considered. As one can expect from our discussion there, $\Delta E$ depends on the magnetic quantum numbers of the molecules. The qualitative behavior of
the effective interaction is the same as in the case of the Born-Oppenheimer approximation, that is, the state $|1010 \rangle$ has the largest effective interaction, followed by $|1000 \rangle$ and $|0000 \rangle$, and the effective interaction is the smallest in case of $|1100 \rangle$. As above, we labeled states according to the first term in Eq.~\eqref{eq:chevy}. In particular, states with $M_1 = 0 = M_2$ have larger effective interaction than states with $M_1, M_2 \neq 0$. The intuition behind this has been explained in detail in Section~\ref{subsec:IIIa}. In contrast to the strong coupling case, the effective interaction does not go to zero for large molecule-molecule distances. As we will see below, this is due to the fact that one phonon cannot dress two impurities in the same way as one phonon dresses a single impurity. 

To investigate this in some more detail, we have a closer look at the self-energy $\Sigma^{\text{BA}}_{L_1M_1L_2M_2}(E_{\text{BA}})$ in Eq. \eqref{eq:SigmaBA} in the limit $d\rightarrow\infty$. The first two terms in this equation are the self-energy contributions of the two molecules, while the third term is related to the effective interaction between them. Since this last term vanishes for $d\rightarrow\infty$, we only need to consider the first two terms. To keep things simple, we also assume that the two molecules are in the same angular momentum state, i.e., $L_1=L_2=l$ and $M_1=M_2=m$. The self-consistent equation (\ref{eq:dyson}) for the energy thus reads
\begin{align}
    &\widetilde{E}_{\text{BA}}(U_\lambda) = 2 Bl(l+1)
    \label{eq:dyson2} \\
    &-\sum_{k\lambda l'}\frac{2\lambda+1}{4\pi}\frac{ 2 U^2_\lambda(k)\left[C^{l'0}_{l0,\lambda0}\right]^2}{Bl'(l'+1)+Bl(l+1)+\omega(k)- \widetilde{E}_{BA}(U_\lambda)},
    \nonumber
\end{align}
where $\widetilde{E}_{\text{BA}}(U_\lambda)= \lim_{d \to \infty} E_{\text{BA}}(U_\lambda)$. We want to compare the solution of this equation to the energy of two separate molecules, that is, to twice the energy of one molecule dressed by one phonon. Such a system has been considered in Ref.~\onlinecite{schmidt2015rotation} and the self-consistent equation for the energy is given by
\begin{align}
    E_{\text{A}}(U_\lambda) =& Bl(l+1) \label{eq:dyson3} \\
    &-\sum_{k\lambda l'}\frac{2\lambda+1}{4\pi}\frac{U^2_\lambda(k)[C^{l'0}_{l0,\lambda0}]^2}{Bl'(l'+1)+\omega(k)-E_{\text{A}}(U_\lambda)} \nonumber
\end{align}
in this case. One easily checks that a solution of \eqref{eq:dyson2} can be written in terms of a solution of \eqref{eq:dyson3} as
\begin{equation}
    \widetilde{E}_{\text{BA}}(U_\lambda) = Bl(l+1)+E_{\text{A}}(\sqrt{2}U_\lambda).
    \label{eq:solution1}
\end{equation}
Here $E_{\text{A}}(\sqrt{2}U_\lambda)$ is the energy of one single molecule but with interaction potential $\sqrt{2}U_\lambda$ instead of $U_\lambda$ in the relevant Hamiltonian. One also checks that the right-hand side of Eq.~\eqref{eq:solution1} is strictly larger than $2 E_{\text{A}}(U_\lambda)$. These results can be explained with the following simple physical picture: The phonon in the system is located with probability $1/2$ close to one molecule and with probability $1/2$ close to the other molecule. This results in an effective potential, which is, compared to the case of one molecule and one phonon, reduced by a factor of $1/\sqrt{2}$ coming from the phonon wave function. The fact that we have a linear coupling and that there are two such interaction terms, one for each molecule, explains the factor of $\sqrt{2} = 2 / \sqrt{2}$ in front of the interaction potential.

The above physical picture is also present in the wave function of the system. If we substitute the relation between the variational coefficients
\begin{align}
&-\beta_{j_1m_1j_2m_2\mathbf{k}}/g =
\nonumber\\
&\hspace{1cm} \frac{e^{-i\frac{1}{2}\mathbf{k}\cdot\mathbf{d}}\langle j_1m_1|\hat{V}|L_1M_1\rangle\delta_{j_2L_2}\delta_{m_2M_2}}{Bj_1(j_1+1)+Bj_2(j_2+1)+\omega(k)-E_{\text{BA}}}
\nonumber\\
&\hspace{1cm}+\frac{e^{i\frac{1}{2}\mathbf{k}\cdot\mathbf{d}}\langle j_2m_2|\hat{V}|L_2M_2\rangle\delta_{j_1L_1}\delta_{m_1M_1}}{Bj_1(j_1+1)+Bj_2(j_2+1)+\omega(k)-E_{\text{BA}}},
\label{eq:galpha}
\end{align}
which follows from the first variation of the energy, into the ansatz Eq. (\ref{eq:chevy}), we find
\begin{eqnarray}
    | \psi_\text{c}\rangle =\frac{1}{\sqrt{2}}\Big[ |L_1M_1\rangle&\otimes&|\psi^{\text{A}}_{L_2M_2} (-d) \rangle
    \nonumber\\
    &+&|\psi^{\text{A}}_{L_1M_1}(d) \rangle \otimes |L_2M_2\rangle \Big].
\label{eq:entangle}
\end{eqnarray}
Here $|\psi^{\text{A}}_{LM} \rangle$ denotes the wave function of one single angulon and reads
\begin{equation}
|\psi^{\text{A}}_{LM} (d) \rangle = \frac{g}{\sqrt{2}}|LM\rangle|0\rangle + \frac{g}{\sqrt{2}}\sum_{j_1\mathbf{k}}f_{L_1,j_1,L_2}(\mathbf{k},d)|j_1m_1\rangle\hat{b}^\dag_{\mathbf{k}}|0\rangle,
\label{eq:angulon_onephonon}
\end{equation}
with
\begin{equation}
f_{l_1,l_2,l_3}(\mathbf{k},d) = \frac{2e^{i\frac{1}{2}\mathbf{k}\cdot\mathbf{d}}\langle l_2|\hat{V}|l_1\rangle}{Bl_3(l_3+1)+Bl_2(l_2+1)+\omega(k)-E_{\text{BA}}}.
\label{eq:angulon_f}
\end{equation}
The wave function of the two impurities in Eq.~\eqref{eq:entangle} is given by an equal weight superposition   of a tensor product of one dressed and one bare molecule, that is, the phonon is with probability $1/2$ located close to the first molecule and with probability $1/2$ close to the second.

From this simple example we learn that one phonon cannot dress each of the two molecules in the same way as one phonon would dress one single molecule. Accordingly, the effective interaction $\Delta E$ \eqref{eq:effectinteraction1ph} does not go to zero as $d \to \infty$, see Fig.~\ref{fig:aainter_chevy}. We checked that this is still true if we consider a trial state with two phonons of the form 
\begin{align}
	|\psi\rangle =& g|L_1M_1\rangle|L_2M_2\rangle|0\rangle + \sum\beta|j_1m_1\rangle|j_2m_2\rangle\hat{b}^\dag_{\mathbf{k}}|0\rangle \\
	&+\sum\gamma|j'_1m'_1\rangle|j'_2m'_2\rangle\hat{b}^\dag_{\mathbf{k}_1}\hat{b}^\dag_{\mathbf{k}_2}|0\rangle, \nonumber
\end{align}
with variational coefficients $g$, $\beta$ and $\gamma$, to compute $E_{\text{BA}}$ (and a trial state with one phonon (or with two phonons) to compute $E_{\text{A},1}$ and $E_{\text{A},2}$). That is, as the above physical picture suggests, two phonons do not dress each of the two molecules (for $d \to \infty$) as one phonon dresses (or two phonons dress) a single impurity. In order to obtain an effective potential with the property $\lim_{d \to \infty} \Delta E = 0$ one would need to consider a sufficiently large number of phonons to compute $E_{\text{BA}}$. In case of a one-phonon or a two-phonon variational state
\begin{equation}
	\Delta E = E_{\mathrm{BA}} - \lim_{d \to \infty} E_{\mathrm{BA}} 
\end{equation}
is therefore clearly a better definition for the effective interaction between the two impurities than Eq.~\eqref{eq:effectinteraction1ph}. Based on the above analysis, we expect that a trial state with one or two phonons yields a good approximation if the distance $d$ between the two impurities is not too large.

\section{\label{sec:level5}The angulon--biangulon transition}
\begin{figure}[t]
	\centering
	\includegraphics[width=0.5\textwidth]{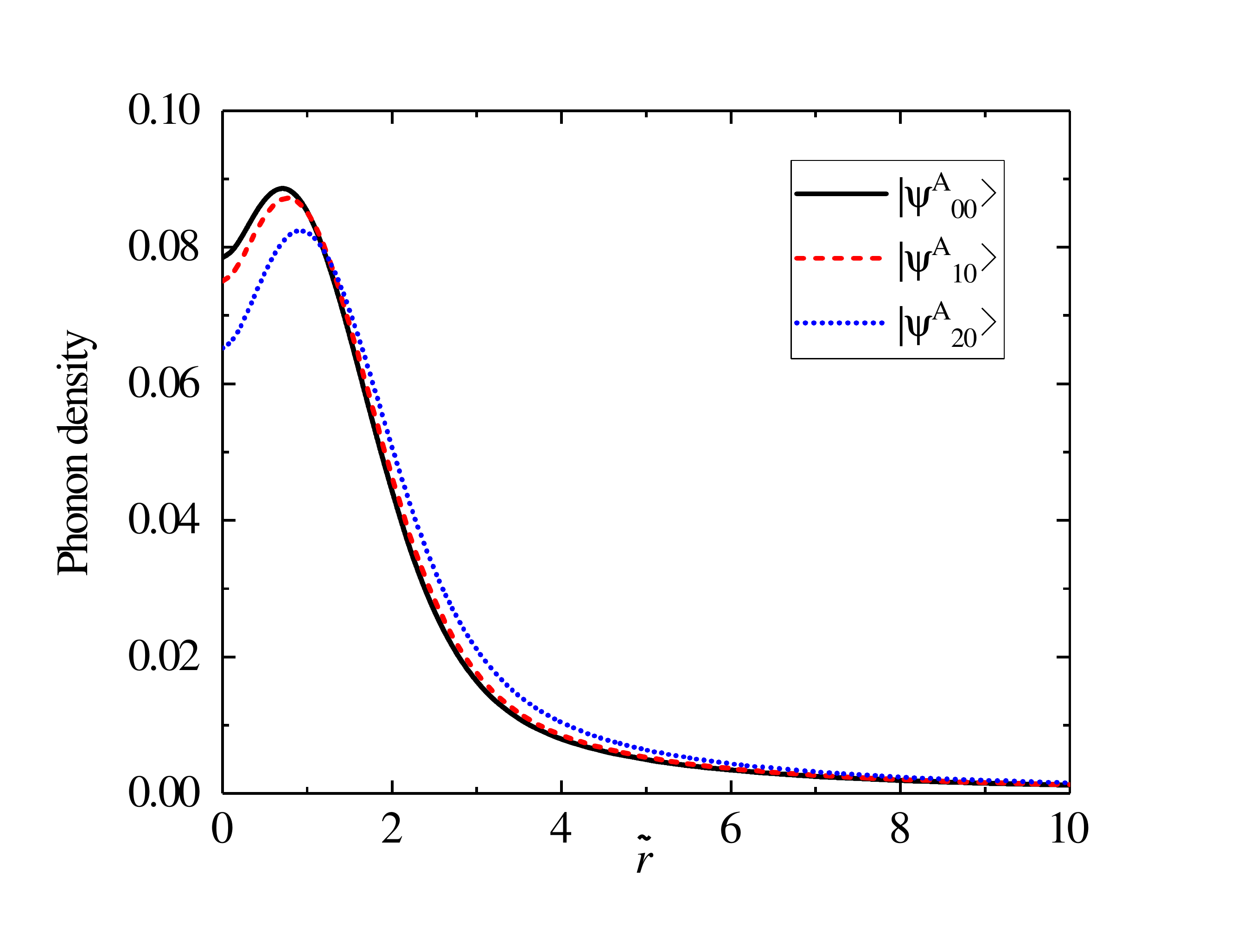}
	\caption{Angle-averaged phonon density $\rho_{LM}(r)$ \eqref{eq:phonondensity} around one single molecule sitting at $\mathbf{r} = 0$ as a function of the dimensionless distance $\tilde{r}=r(mB)^{-1/2}$ to the origin. We have chosen $u_0=u_1=u_2=218 B$, $u_{\lambda} = 0$ for $\lambda \geq 3$ and $\tilde{r}_0=\tilde{r}_1=\tilde{r}_2=1.5(mB)^{-1/2}$, $\tilde{r}_\lambda = 0$ if $\lambda \geq 3$ as well as $\tilde{n}=1$. The quantum numbers of the angulon are $L=0, M=0$ (solid black line), $L=1, M=0$ (red dashed line), $L=2, M=0$ (blue dotted line). For more information see the text.}
	\label{fig:density}
\end{figure}
As explained in detail in Section~\ref{sec:ang}, the biangulon quasi-article is defined by a strongly correlated relative alignment of the two molecules. Additionally, the phonon cloud related to one molecule has a substantial overlap with the other molecule (and the other way round). The magnetic quantum number $M$ of the total system is the biangulon's only good quantum number. The term related to the quasi particle weight $g$ in Eq.~\eqref{eq:chevy} has quantum numbers $L_1,M_1,L_2,M_2$. Accordingly, the trial state is well suited to either describe a moderately correlated biangulon or two weakly interacting angulons. A more general biangulon state should allow for a substantial mixing of the different basis states $| L_1,M_1,L_2,M_2 \rangle$ with $M_1+M_2 = M$ in the term proportional to the quasiparticle weight. In this Section we are going to study the transition from two weakly interacting angulons to a strongly correlated biangulon, and therefore choose a different approach than in Section~\ref{sec:level4} that takes the above consideration into account.

In order to simplify the analysis, we consider a situation, where the interaction of one of the impurities with the bath is weaker than that of the other impurity. This could correspond e.g. to the physical picture of one heavier and one lighter molecule. The system will be described by wave functions of the form
\begin{equation}
|\psi_d\rangle = \sum_{L,M,j,m}\alpha^{L,M}_{j,m}|\psi^{\text{A}}_{L,M}\rangle|jm\rangle.
\label{eq:base}
\end{equation}
Here $\alpha^{L,M}_{j,m}$ are variational coefficients that obey the usual normalization condition and assure that $M + m = \widetilde{M}$ holds with some fixed $\widetilde{M}$. Additionally,
\begin{align}
|\psi^{\text{A}}_{LM}\rangle=&\sqrt{Z_{L}}|LM\rangle|0\rangle
\nonumber\\
&+\sum_{k\lambda j_1}\beta_{k\lambda j_1}C^{LM}_{j_1m_1,\lambda\mu}|j_1m_1\rangle\hat{b}^\dag_{k\lambda\mu}|0\rangle
\label{eq:angulon}  
\end{align}
denotes the wave function of one single angulon with angular momentum quantum numbers $L, M$. We obtain the coefficients in Eq. \eqref{eq:angulon} by considering the relevant one-impurity system, see Ref.~\onlinecite{schmidt2015rotation}. The impurity described by the first tensor factor in Eq.  \eqref{eq:base} is the one with stronger molecule-bath interaction, and therefore it is assumed to be already dressed by the phonon in the system. The second impurity is described by a free rotor. Due to the generality of the variational coefficients, the above ansatz allows for a substantial mixing of different free rotor states in the part of the wave function with no phonons. Using it, we can therefore describe the transition from two weakly coupled angulons, where the wave function is approximately given by $|\psi^{\text{A}}_{L,M}\rangle|jm\rangle$ for some quantum numbers $L,M,j,m$, to a strongly correlated biangulon quasiparticle, where more than one of the coefficients $\alpha^{L,M}_{j,m}$ are unequal to zero.  
The above ansatz efficiently describes phonon-induced interactions between the two molecules as long as the weakly interacting impurity has a substantial overlap with the phonon density located around the first molecule. 

In Fig.~\ref{fig:density} we show an example of such a phonon density. More precisely, we show the angle-averaged phonon density 
\begin{align}
		\rho_{LM}(r) &= \int  \mathrm{d} \phi_{\mathbf{r}} \mathrm{d} \theta_{\mathbf{r}} \langle \psi^{\text{A}}_{L,M}|\hat{b}^\dag_{\mathbf{r}}\hat{b}_{\mathbf{r}}|\psi^{\text{A}}_{L,M}\rangle  \nonumber \\
		&= \sum_{\lambda\mu}\langle\psi^{\text{A}}_{L,M}|\hat{b}^\dag_{r\lambda\mu}\hat{b}_{r\lambda\mu}|\psi^{\text{A}}_{L,M}\rangle 
		\label{eq:phonondensity}
\end{align}
of one single impurity described by the angulon wave function \eqref{eq:angulon}. Here $\hat{b}^\dag_{\mathbf{r}}$ creates one phonon at position $\mathbf{r}$ and we used
\begin{align}
	\hat{b}^\dag_{\mathbf{r}} &= \frac{1}{r} \sum_{\lambda \mu} \hat{b}^\dag_{r\lambda\mu} Y_{\lambda \mu}^*(\theta_{\mathrm{r}},\phi_{\mathrm{r}}),
\end{align}
see \onlinecite{lemeshko2017molecular}. The operator $\hat{b}^\dag_{r\lambda\mu}$ creates one phonon at distance $r$ from the origin with angular momentum quantum numbers $\lambda, \mu$. It can be written in terms of the operators $\hat{b}^\dag_{k\lambda\mu}$ as
\begin{equation}
	\hat{b}^\dag_{r\lambda\mu} = \sqrt{\frac{2}{\pi} } r \int k \mathrm{d} k j_{\lambda}(kr) \hat{b}^\dag_{k\lambda\mu}, 
\end{equation}
where $j_{\lambda}(kr)$ denotes the spherical Bessel function \cite{abramowitz1965handbook}. The parameters are chosen to be $u_0=u_1=u_2=218 B$, $u_{\lambda} = 0$ for $\lambda \geq 3$ and $\tilde{r}_0=\tilde{r}_1=\tilde{r}_2=1.5(mB)^{-1/2}$, $\tilde{r}_\lambda = 0$ if $\lambda \geq 3$. The density is given by $\tilde{n}=1$ and the quantum numbers of the angulon are chosen as $L=0, M=0$ (solid black line), $L=1, M=0$ (red dashed line), $L=2, M=0$ (blue dotted line). As long as the distance between the two impurities is below $\tilde{d} \approx 6$ for this choice of the parameters, the ansatz \eqref{eq:base} allows us to capture the interactions between the two impurities.
	
For mathematical convenience we assume from now on that the stronger interacting impurity is sitting at the origin of the laboratory frame and that the weaker interacting impurity is located at $(0,0,d)$. To diagonalize the biangulon Hamiltonian \eqref{eq:hbiang2} with the basis set \eqref{eq:base}, we write it as $\hat{H} = \hat{H}_\text{A} + \hat{H}_{\text{I}}$, where
\begin{align}
	\hat{H}_{\text{A}} =& B_1\hat{\mathbf{J}}^2_1 + B_2\hat{\mathbf{J}}^2_2 + \sum_{\mathbf{k}}\omega(k)\hat{b}^\dag_{\mathbf{k}}\hat{b}_{\mathbf{k}} \nonumber \\
	&\hspace{2cm}+ \sum_{k\lambda\mu}\left[V(\mathbf{k},\hat{\theta}_1,\hat{\phi}_1) \hat{b}^\dag_{\mathbf{k}}+\text{H.c.}\right]
	\label{eq:hbiang3a}
\end{align}
	and
	\begin{equation}
	\hat{H}_{\text{I}} = \sum_{\mathbf{k}}\left[V(\mathbf{k},\hat{\theta}_2,\hat{\phi}_2)e^{i\mathbf{k}\cdot\mathbf{d}}\hat{b}^\dag_{\mathbf{k}} +  \text{H.c.}\right] \; .
	\label{eq:hbiang3b} 
	\end{equation}
	The Hamiltonian $\hat{H}_{\text{A}}$ describes a single angulon \cite{schmidt2015rotation,lemeshko2017quasiparticle} and a bare rotating molecule, and can therefore be considered as diagonal within our approximation scheme. This allows us to write the matrix elements of the biangulon Hamiltonian $\hat{H}$ with respect to the basis states in Eq.~\eqref{eq:base} as
	\begin{eqnarray}
	H^{L'M'j'm'}_{LMjm} &=& \left[E^{L,M}_{\text{A}}+Bj(j+1)\right]\delta_{L',L}\delta_{M',M}\delta_{j',j}\delta_{m',m}
	\nonumber\\
	& &+ \langle\psi^{\text{A}}_{L',M'}|\langle j'm'|\hat{H}_{\text{I}}|jm\rangle|\psi^{\text{A}}_{L,M}\rangle.
	\label{eq:element}
	\end{eqnarray}
	In order to obtain the energies and eigenfunctions, we diagonalize the Hamiltonian matrix \eqref{eq:element} numerically with the angular momentum cut-off $L,L',j,j', |M|,|M'|, |m|,|m'| \leq 2$.
	
\begin{figure}[t]
	\centering
	\includegraphics[width=0.5\textwidth]{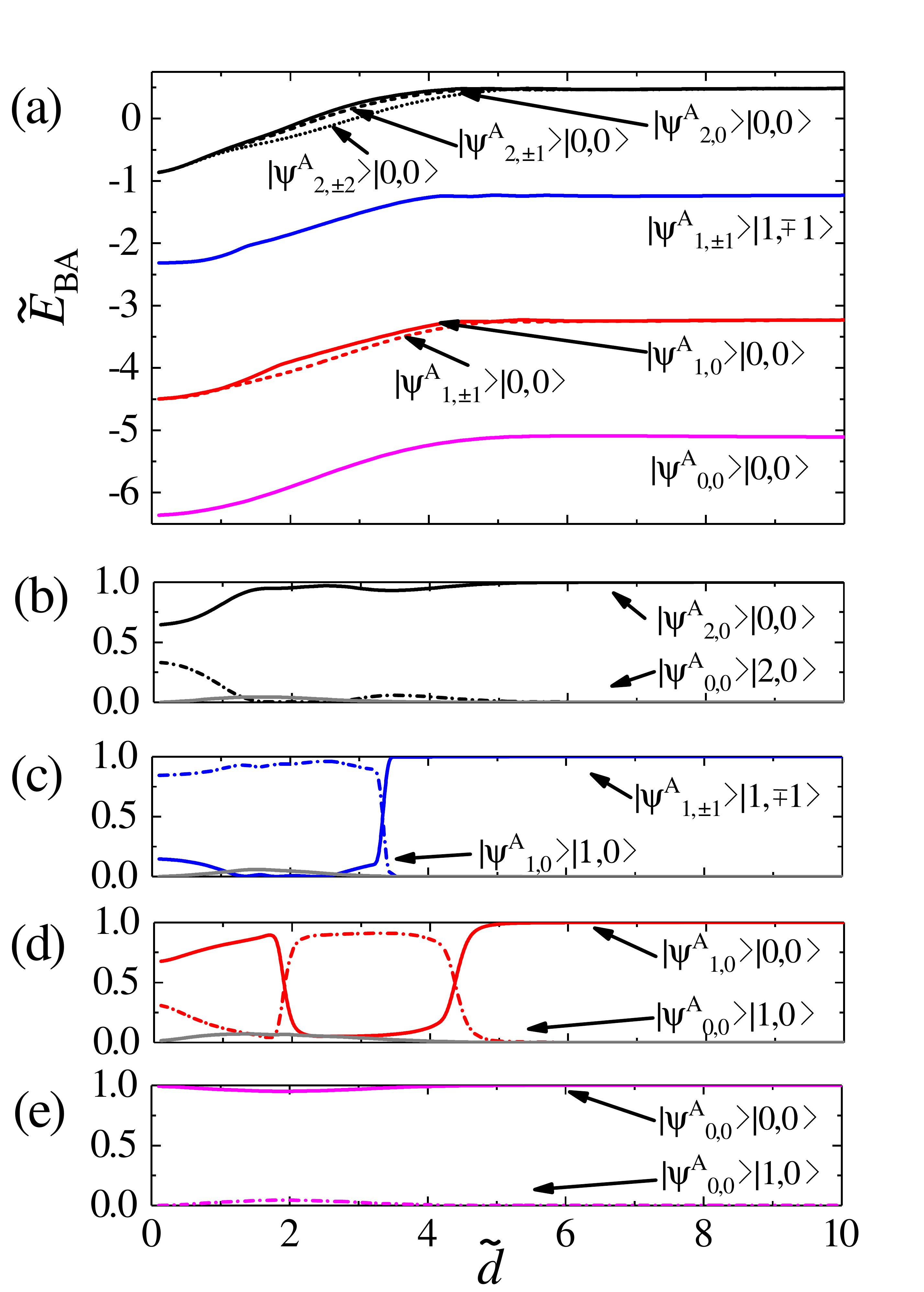}
	\caption{(a) The dimensionless biangulon energy of the ground state and of six excited states obtained by diagonalizing the biangulon Hamiltonian \eqref{eq:hbiang2} with the base vectors used in \eqref{eq:base}. In (b)--(e) we show the squared overlap of the eigenstate $|\psi^{\text{A}}_{2,0} ; 0,0 \rangle$ (b), $|\psi^{\text{A}}_{1,\pm 1} ; 1, \mp 1 \rangle$ (c), $|\psi^{\text{A}}_{1,0} ; 1, 0 \rangle$ (d) and $|\psi^{\text{A}}_{0,0} ; 0, 0 \rangle$ (e) with the different basis states. The bath density has been chosen as $\tilde{n}=1$. The grey lines show the occupation all other basis vectors. For more information see the text.}
	\label{fig:eigen}
\end{figure}

As parameters we choose $u_{\lambda,1} = 2 u_{\lambda,2}$, where the second index refers to the first and the second impurity, $u_{0,1}=u_{1,1}=u_{2,1}=218 B$ and $\tilde{n} = 1$. We label eigenstates by their dominant basis vector contribution at $\tilde{d} = 10$, that is, at that distance the eigenfunction $|\psi^{\text{A}}_{L,M}; j,m \rangle$ approximately equals $|\psi^{\text{A}}_{L,M} \rangle | j,m \rangle$. The results of the diagonalization are presented in Fig.~\ref{fig:eigen}. In Fig.~\ref{fig:eigen}(a) we show the energy of the ground state $|\psi^{\text{A}}_{0,0}; 0,0\rangle$ and of six excited states. States which differ only by the magnetic quantum number of the two molecules are degenerate if the distance between them is sufficiently large because $E^{L,M}_{\text{A}} = E^{L,-M}_{\text{A}}$. This degeneracy is lifted when the particles start to substantially interact around $\tilde{d} \approx 6$. In this regime the eigenvalues related to $|\psi^{\text{A}}_{1,0} ; 0,0\rangle$ (red solid line) and $|\psi^{\text{A}}_{2,0} ; 0,0\rangle$ (solid black line) start to split from those related to $|\psi^{\text{A}}_{1,\pm 1} ; 0,0\rangle$ (red dashed line) and $|\psi^{\text{A}}_{2,\pm 1} ;0,0\rangle$ (black dashed line), $|\psi^{\text{A}}_{2,\pm 2}; 0,0\rangle$ (black dotted line), respectively. The states $|\psi^{\text{A}}_{1,\pm 1} ;1,\mp 1 \rangle$ remain degenerate. 

In Fig.~\ref{fig:eigen} In (b)--(e) we show the squared overlap of the eigenstate $|\psi^{\text{A}}_{2,0} ; 0,0 \rangle$ (b), $|\psi^{\text{A}}_{1,\pm 1} ; 1, \mp 1 \rangle$ (c), $|\psi^{\text{A}}_{1,0} ; 1, 0 \rangle$ (d) and $|\psi^{\text{A}}_{0,0} ; 0, 0 \rangle$ (e) with the different basis states. We note that all these states have $M + m = 0$. The grey lines show the occupation of all other basis vectors. As can be seen from these figures, different eigenstates of the Hamiltonian matrix \eqref{eq:element} show different behavior during the transition from two separate angulons to a biangulon if the distance between them is decreased. The states $|\psi^{\text{A}}_{1,\pm 1} ; 1,\mp 1\rangle$ and $|\psi^{\text{A}}_{1,0} ; 0,0 \rangle$ for example show a sharp transition, while this transition is less pronounced for the state $|\psi^{\text{A}}_{2,0} ; 0,0 \rangle$ and it is almost not present in case of the ground state $|\psi^{\text{A}}_{0,0 } ; 0, 0\rangle$. This behavior is a result of the SO($3$) algebra of angular momentum ruling the interaction between the two impurities. In general, we can say that the states with $M=0=m$ and $L \neq j$ show the most pronounced angulon to biangulon transitions. In case of $M=0=m$ the wave function is with good approximation a superposition of two basis states. As an example we consider states of the form
\begin{equation}\label{eq:2}
	|\psi^{\text{A}}_{L,0}; j,0\rangle \approx c_1(d) |\psi^{\text{A}}_{L,0}\rangle|j,0\rangle + c_2(d) |\psi^{\text{A}}_{j,0}\rangle|L,0\rangle,
\end{equation}
compare with Fig.~\ref{fig:eigen}(b) and (d). This representation implies that angular momentum is transferred from one impurity to the other during the transition from two separated angulons to a biangulon quasiparticle. The fact that exactly these two basis states appear in Eq.~\eqref{eq:2} is again a result of the SO($3$) algebra of angular momentum. For several other basis states we find a similar but less pronounced angulon-biangulon transition. The weakest transition can be seen in states of the form $|\psi^{\text{A}}_{L,0} ; L, 0 \rangle$.

In order to investigate the transition from two angulons to a biangulon for states that show a pronounced transition in more detail, we consider correlation functions of the form
\begin{equation}\label{eq:3}
	F_{\hat{O}} = \frac{ \langle \hat{O}_1 \hat{O}_2 \rangle - \langle \hat{O}_1 \rangle \langle \hat{O}_2 \rangle }{\langle \hat{O}_1 \hat{O}_2 \rangle_{\mathrm{max}} - \langle \hat{O}_1 \rangle_{\mathrm{max}} \langle \hat{O}_2 \rangle_{\mathrm{max}} },
\end{equation}
where $\langle \cdot \rangle$ denotes the expectation w.r.t. one of the eigenfunctions of the two impurity problem and $\hat{O}_i$, $i=1,2$, is an operator acting on the $i$-th impurity. As an example, we consider eigenstates that can with a good approximation be written as a distance-dependent superposition of two basis states $|v \rangle$ and $|w \rangle$, that is, states of the form
\begin{equation}\label{eq:4}
	| \psi_{d} \rangle \approx c_1(d) | v \rangle + c_2(d) | w \rangle,
\end{equation}
compare with Eq.~\eqref{eq:2}. The normalization in \eqref{eq:3} is chosen such that $|F_{\hat{O}}|$ takes values between zero and one. More precisely, we assume that the expectation $\langle \cdot \rangle_{\mathrm{max}}$ is taken with respect to the state
\begin{equation}
	| \psi_{\mathrm{max}} \rangle = \frac{1}{\sqrt{2}} \left( | v \rangle + | w \rangle \right).
\end{equation}
In the cases we consider, the state $| \psi_{\mathrm{max}} \rangle$ maximizes the correlation function amoung normalized states of the form given by Eq.~\eqref{eq:4}. Since the different eigenfunctions of the Hamiltonian matrix \eqref{eq:element} we consider here have different dominant basis vectors in their expansion we also have to use different operators $\hat{O}$ to measure their correlations.

\begin{figure}[t]
	\centering
	\includegraphics[width=0.5\textwidth]{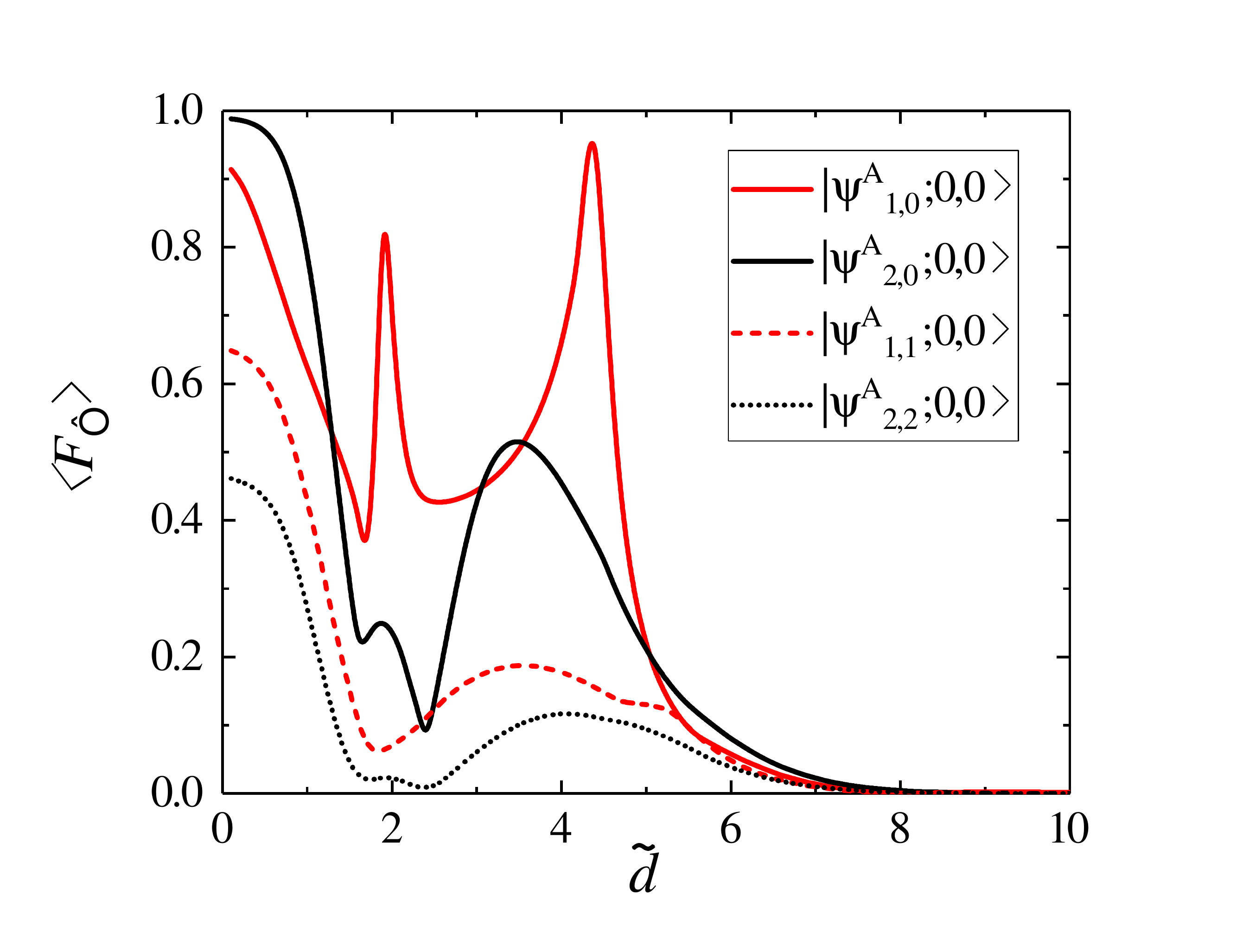}
	\caption{Correlation function $F_{\hat{O}}$, Eq.~\eqref{eq:3}, as a function of the dimensionless  molecule-molecule distance $\tilde{d}$. The parameters are the same as in Fig.~\ref{fig:eigen}. The colors of the graphs refer to the same states as in Fig.~\ref{fig:eigen}(a). For the operator $\hat{O}$ and for the state $\langle \cdot \rangle$ we made the following choice: $\hat{O}= \cos( \theta )$, $|\psi^{\text{A}}_{1,0}; 0,0 \rangle $ (red solid line), $\hat{O} = \cos^2( \theta ) $, $|\psi^{\text{A}}_{2,0}; 0,0 \rangle $ (solid black line), $\hat{O} = \sin(\theta)e^{\pm i\varphi}$, $|\psi^{\text{A}}_{1,1}; 0,0 \rangle $ (red dashed line) and $\hat{O}= \sin^2(\theta)e^{\pm i2 \varphi} $, $|\psi^{\text{A}}_{2,2}; 0,0 \rangle $ (black dotted line). For more information see the text.}
	\label{fig:correlation}
\end{figure}

The correlation functions related to four eigenstates of the Hamiltonian matrix can be found in Fig.~\ref{fig:correlation}. We have chosen $\hat{O}= \cos( \theta )$, $|\psi^{\text{A}}_{1,0}; 0,0 \rangle $ (red solid line), $\hat{O} = \cos^2( \theta ) $, $|\psi^{\text{A}}_{2,0}; 0,0 \rangle $ (solid black line), $\hat{O} = \sin(\theta)e^{\pm i\varphi}$, $|\psi^{\text{A}}_{1,1}; 0,0 \rangle $ (red dashed line) and $\hat{O}= \sin^2(\theta)e^{\pm i2 \varphi} $, $|\psi^{\text{A}}_{2,2}; 0,0 \rangle $ (black dotted line). The interaction between the impurities is attractive, and hence all correlation functions are positive. The particular patterns that these functions show are related to the shape of our interaction potential. All correlation functions indicate that after the onset of interactions between the two impurities around $\tilde{d} \sim 6$, the eigenstates of the Hamiltonian matrix \eqref{eq:element} we considered in Fig.~\ref{fig:correlation} quickly start to be substantially entangled and correlated when the distance between them is further reduced -- a clear signature that a biangulon quasiparticle forms. 

A similar but less pronounced behavior can be found for several other eigenstates. The states $|\psi^{\text{A}}_{L,0} ; L, 0 \rangle$ show, however, almost no correlations and have $|\psi^{\text{A}}_{L,0} \rangle | L, 0 \rangle$ as a dominant basis vector for all distances. The weakest correlation can be found in the ground state. The fact that its wave function is with good approximation given by $|\psi^{\text{A}}_{0,0} \rangle | 0, 0 \rangle$ is in accordance with the analysis in the strong-coupling regime in Sec.~\ref{subsec:IIIb}, where we found that the ground state is a product of two (the same) impurity wave function. Here the system looked like a biangulon quasiparticle because of the substantial anisotropy of the molecular orientations and because the phonon cloud related to one molecules had a substantial overlap with the other molecule (and the other way round). Due to the simplicity of our approach, this is clearly not captured by the analysis in this Section. To take such effects into account, which would allow us to investigate the transition from two separate angulons to a biangulon also for the states $|\psi^{\text{A}}_{L,0} ; L, 0 \rangle$ in more detail, we would need to allow for more basis states in the expansion of the molecular states. Additionally, we would need to treat also the phonon wave function variationally. This, however, is beyond the scope of the present paper. 

\section{\label{sec:level6}Conclusion}

By applying translation operators to the previously introduced angulon Hamiltonian, we obtained the Hamiltonian describing two rotating molecules immersed in a bosonic bath. This model was studied in different parameter regimes and using several theoretical approaches. In all the parameter regimes we found that the molecules align with respect to each other as a result of the phonon mediated effective attractive interaction~\eqref{eq:VAA} between them. To describe the resulting correlated state, we introduced the \emph{biangulon} quasiparticle. In analogy to the bipolaron quasiparticle, it describes two rotating molecules dressed by bosonic excitations. 

We first considered the regime where the molecular rotation is much slower than the characteristic timescale of the phonons. In this situation the phonon cloud adjusts itself instantaneously to changes of the molecular orientation and a Born-Oppenheimer approximation is valid. Within this approach we showed that the effective intermolecular force mediated by the phonons is sensitive to the rotational state of both molecules and takes its largest values when the overlap of the phonon density with each of the two molecules is maximal. Accordingly, the states with magnetic quantum numbers $M_1=0=M_2$, which preserve the symmetries of the Hamiltonian, show the largest effective interaction.

The ground state of the system in the strong-coupling regime has been investigated by minimizing the related Pekar functional. In this model the two molecules co-align in order to maximize the overlap with the phonon cloud. As a consequence, the presence of the bath can enhance the rate of certain chemical reactions that favour such an alignment between the molecules.

In the opposite regime, where the impurity-bath coupling is relatively weak, we investigated the system with the help of a one-phonon excitation variational ansatz, which allowed us to access the excitation spectrum of the biangulon. In comparison to the angulon spectrum, we observed an additional spectral instability, where a resonant angular momentum transfer between molecules and the bath takes place, as well as a shift of the angulon spectral instabilities due to the presence of the second molecule. These features have been proposed as a signature for the formation of the biangulon quasiparticle in experiments. Additionally, we pointed out that in our model one or two phonons cannot dress two molecules that are far apart from each other as one phonon dresses one single molecule, which leads to a subtlety in the definition of the effective phonon-mediated interaction for large distances.

Finally, by using products of angulon and bare rotor states as basis states, we investigated the system in the situation where the interaction of one of the two impurities with the bath is substantially weaker than that of the other. This approach allowed us to study the transition from two separated angulons to a biangulon quasiparticle as a function of the distance between the two molecules. In the parameter regime where a biangulon has formed, the wavefunction is a superposition of at least two of the above basis states. Accordingly, angular momentum is transferred between the two molecules and the state is strongly correlated. This has to be contrasted with the appearance of two uncorrelated or weakly correlated angulons at larger molecule-molecule distance. 

The above results can be applied to molecules immersed in superfluid helium droplets\cite{Toennies2004} or in atomic Bose-Einstein condensates\cite{lemeshko2017molecular}, and can be extended to systems where the impurity particles are Rydberg atoms\cite{camargo2016creation,camargo2018creation} or defects in solids\cite{pushkarov1991quasiparticle}.

 \section{\label{sec:level7} acknowledgements} 
 We are grateful to Areg Ghazaryan for valuable discussions. M.L.~acknowledges support by the Austrian Science Fund (FWF), under project No.~P29902-N27, and by the European Research Council (ERC) Starting Grant No.~801770 (ANGULON). E.Y. acknowledges financial support received from the People Programme (Marie Curie Actions) of the European Union's Seventh Framework Programme (FP7/2007-2013) under REA grant agreement No.~[291734]. G.B.~acknowledges support from the Austrian Science Fund (FWF), under project No.~M2461-N27. A.D. acknowledges funding from the European Union’s Horizon 2020 research and innovation programme under the European Research Council (ERC) grant agreement No~694227 and under the Marie Sklodowska-Curie grant agreement No~836146. R.S. is supported by the Deutsche Forschungsgemeinschaft (DFG, German Research Foundation) under Germany's Excellence Strategy -- EXC-2111 -- 390814868.
\appendix

\section{\label{sec:appendix}Simulated annealing}
The stochastic simulated annealing method that we applied to minimize ground energy Eq. \ref{eq:pekar_EBA}, is based on repeated application of the following two moves: 

\begin{enumerate}
\item Rotation of a variational coefficient in the complex plane, i.e.~$s_{c} \to s_{c} \exp(\mathrm{i} \phi)$ where the quantum numbers $c=(L_1, M_1, L_2, M_2)$ and the phase $\phi$ have been chosen from a random distribution.
\item Moving part of the complex modulus of a coefficient to another coefficient, i.e.~going from a configuration of two coefficients that we parametrize in the polar representation as $s_{c} = \rho \exp(\mathrm{i} \phi)$, $s_{c'} = (\rho') \exp(\mathrm{i} \phi')$ to a different configuration $s_{c} = (\rho - \delta) \exp(\mathrm{i} \phi)$, $s_{c'} = (\rho' + \delta) \exp(\mathrm{i} \phi')$ where again the quantum numbers $c$ and $c'$, as well as $\delta$, are chosen randomly.
\end{enumerate}

It can be easily seen that these two moves span the whole parameter space, while automatically enforcing the normalization condition. In the spirit of simulated annealing methods, each move is accepted or rejected by evaluating the Boltzmann factor of the energy difference, using a monotonously decreasing effective temperature. We have verified that this procedure is solid, yielding a good estimate of the ground state energy at the level of maximum $L_i=4$ (containing $1,764$ variational coefficients), independently of the starting configuration, in agreement with non-stochastic methods that are  usually slower and limited to much smaller cutoffs.

\vspace{1cm}
\bibliography{biangulon}

\begin{thebibliography}{60}%
\makeatletter
\providecommand \@ifxundefined [1]{%
 \@ifx{#1\undefined}
}%
\providecommand \@ifnum [1]{%
 \ifnum #1\expandafter \@firstoftwo
 \else \expandafter \@secondoftwo
 \fi
}%
\providecommand \@ifx [1]{%
 \ifx #1\expandafter \@firstoftwo
 \else \expandafter \@secondoftwo
 \fi
}%
\providecommand \natexlab [1]{#1}%
\providecommand \enquote  [1]{``#1''}%
\providecommand \bibnamefont  [1]{#1}%
\providecommand \bibfnamefont [1]{#1}%
\providecommand \citenamefont [1]{#1}%
\providecommand \href@noop [0]{\@secondoftwo}%
\providecommand \href [0]{\begingroup \@sanitize@url \@href}%
\providecommand \@href[1]{\@@startlink{#1}\@@href}%
\providecommand \@@href[1]{\endgroup#1\@@endlink}%
\providecommand \@sanitize@url [0]{\catcode `\\12\catcode `\$12\catcode
  `\&12\catcode `\#12\catcode `\^12\catcode `\_12\catcode `\%12\relax}%
\providecommand \@@startlink[1]{}%
\providecommand \@@endlink[0]{}%
\providecommand \url  [0]{\begingroup\@sanitize@url \@url }%
\providecommand \@url [1]{\endgroup\@href {#1}{\urlprefix }}%
\providecommand \urlprefix  [0]{URL }%
\providecommand \Eprint [0]{\href }%
\providecommand \doibase [0]{http://dx.doi.org/}%
\providecommand \selectlanguage [0]{\@gobble}%
\providecommand \bibinfo  [0]{\@secondoftwo}%
\providecommand \bibfield  [0]{\@secondoftwo}%
\providecommand \translation [1]{[#1]}%
\providecommand \BibitemOpen [0]{}%
\providecommand \bibitemStop [0]{}%
\providecommand \bibitemNoStop [0]{.\EOS\space}%
\providecommand \EOS [0]{\spacefactor3000\relax}%
\providecommand \BibitemShut  [1]{\csname bibitem#1\endcsname}%
\let\auto@bib@innerbib\@empty
\bibitem [{\citenamefont {Devreese}(2015)}]{Devreese15}%
  \BibitemOpen
  \bibfield  {author} {\bibinfo {author} {\bibfnamefont {J.~T.}\ \bibnamefont
  {Devreese}},\ }\bibfield  {title} {\enquote {\bibinfo {title} {Lectures on
  {Fr{\"o}hlich} polarons from $3{D}$ to $0{D}$ --- including detailed
  theoretical derivations},}\ }\href {https://arxiv.org/abs/1012.4576}
  {\bibfield  {journal} {\bibinfo  {journal} {arXiv:1012.4576v6}\ } (\bibinfo
  {year} {2015})}\BibitemShut {NoStop}%
\bibitem [{\citenamefont {Devreese}\ and\ \citenamefont
  {Alexandrov}(2009)}]{devreese2009frohlich}%
  \BibitemOpen
  \bibfield  {author} {\bibinfo {author} {\bibfnamefont {J.~T.}\ \bibnamefont
  {Devreese}}\ and\ \bibinfo {author} {\bibfnamefont {A.~S.}\ \bibnamefont
  {Alexandrov}},\ }\bibfield  {title} {\enquote {\bibinfo {title} {Fr{\"o}hlich
  polaron and bipolaron: recent developments},}\ }\href@noop {} {\bibfield
  {journal} {\bibinfo  {journal} {Reports on Progress in Physics}\ }\textbf
  {\bibinfo {volume} {72}},\ \bibinfo {pages} {066501} (\bibinfo {year}
  {2009})}\BibitemShut {NoStop}%
\bibitem [{\citenamefont {Kashirina}\ and\ \citenamefont
  {Lakhno}(2010)}]{kashirina2010large}%
  \BibitemOpen
  \bibfield  {author} {\bibinfo {author} {\bibfnamefont {N.~I.}\ \bibnamefont
  {Kashirina}}\ and\ \bibinfo {author} {\bibfnamefont {V.~D.}\ \bibnamefont
  {Lakhno}},\ }\bibfield  {title} {\enquote {\bibinfo {title} {Large-radius
  bipolaron and the polaron--polaron interaction},}\ }\href@noop {} {\bibfield
  {journal} {\bibinfo  {journal} {Physics-Uspekhi}\ }\textbf {\bibinfo {volume}
  {53}},\ \bibinfo {pages} {431} (\bibinfo {year} {2010})}\BibitemShut
  {NoStop}%
\bibitem [{\citenamefont {Alexandrov}(2003)}]{alexandrov2003theory}%
  \BibitemOpen
  \bibfield  {author} {\bibinfo {author} {\bibfnamefont {A.~S.}\ \bibnamefont
  {Alexandrov}},\ }\href@noop {} {\emph {\bibinfo {title} {Theory of
  superconductivity: from weak to strong coupling}}}\ (\bibinfo  {publisher}
  {CRC Press},\ \bibinfo {year} {2003})\BibitemShut {NoStop}%
\bibitem [{\citenamefont {Qu{\'e}merais}\ and\ \citenamefont
  {Fratini}(1998)}]{quemerais1998polaron}%
  \BibitemOpen
  \bibfield  {author} {\bibinfo {author} {\bibfnamefont {P.}~\bibnamefont
  {Qu{\'e}merais}}\ and\ \bibinfo {author} {\bibfnamefont {S.}~\bibnamefont
  {Fratini}},\ }\bibfield  {title} {\enquote {\bibinfo {title} {Polaron
  crystallization and the metal--insulator transition},}\ }\href@noop {}
  {\bibfield  {journal} {\bibinfo  {journal} {International Journal of Modern
  Physics B}\ }\textbf {\bibinfo {volume} {12}},\ \bibinfo {pages} {3131--3136}
  (\bibinfo {year} {1998})}\BibitemShut {NoStop}%
\bibitem [{\citenamefont {Fratini}\ and\ \citenamefont
  {Qu{\'e}merais}(2002)}]{fratini2002polarization}%
  \BibitemOpen
  \bibfield  {author} {\bibinfo {author} {\bibfnamefont {S.}~\bibnamefont
  {Fratini}}\ and\ \bibinfo {author} {\bibfnamefont {P.}~\bibnamefont
  {Qu{\'e}merais}},\ }\bibfield  {title} {\enquote {\bibinfo {title}
  {Polarization catastrophe in the polaronic wigner crystal},}\ }\href@noop {}
  {\bibfield  {journal} {\bibinfo  {journal} {The European Physical Journal
  B-Condensed Matter and Complex Systems}\ }\textbf {\bibinfo {volume} {29}},\
  \bibinfo {pages} {41--49} (\bibinfo {year} {2002})}\BibitemShut {NoStop}%
\bibitem [{\citenamefont {Iadonisi}\ \emph {et~al.}(2007)\citenamefont
  {Iadonisi}, \citenamefont {Mukhomorov}, \citenamefont {Cantele},\ and\
  \citenamefont {Ninno}}]{iadonisi2007formation}%
  \BibitemOpen
  \bibfield  {author} {\bibinfo {author} {\bibfnamefont {G.}~\bibnamefont
  {Iadonisi}}, \bibinfo {author} {\bibfnamefont {V.}~\bibnamefont
  {Mukhomorov}}, \bibinfo {author} {\bibfnamefont {G.}~\bibnamefont {Cantele}},
  \ and\ \bibinfo {author} {\bibfnamefont {D.}~\bibnamefont {Ninno}},\
  }\bibfield  {title} {\enquote {\bibinfo {title} {Formation of a large polaron
  crystal from a homogeneous, dilute polaron gas},}\ }\href@noop {} {\bibfield
  {journal} {\bibinfo  {journal} {Physical Review B}\ }\textbf {\bibinfo
  {volume} {76}},\ \bibinfo {pages} {144303} (\bibinfo {year}
  {2007})}\BibitemShut {NoStop}%
\bibitem [{\citenamefont {Kusmartsev}(2001)}]{kusmartsev2001electronic}%
  \BibitemOpen
  \bibfield  {author} {\bibinfo {author} {\bibfnamefont {F.}~\bibnamefont
  {Kusmartsev}},\ }\bibfield  {title} {\enquote {\bibinfo {title} {Electronic
  molecules in solids},}\ }\href@noop {} {\bibfield  {journal} {\bibinfo
  {journal} {EPL (Europhysics Letters)}\ }\textbf {\bibinfo {volume} {54}},\
  \bibinfo {pages} {786} (\bibinfo {year} {2001})}\BibitemShut {NoStop}%
\bibitem [{\citenamefont {Perroni}, \citenamefont {Iadonisi},\ and\
  \citenamefont {Mukhomorov}(2004)}]{perroni2004formation}%
  \BibitemOpen
  \bibfield  {author} {\bibinfo {author} {\bibfnamefont {C.}~\bibnamefont
  {Perroni}}, \bibinfo {author} {\bibfnamefont {G.}~\bibnamefont {Iadonisi}}, \
  and\ \bibinfo {author} {\bibfnamefont {V.}~\bibnamefont {Mukhomorov}},\
  }\bibfield  {title} {\enquote {\bibinfo {title} {Formation of polaron
  clusters},}\ }\href@noop {} {\bibfield  {journal} {\bibinfo  {journal} {The
  European Physical Journal B-Condensed Matter and Complex Systems}\ }\textbf
  {\bibinfo {volume} {41}},\ \bibinfo {pages} {163--170} (\bibinfo {year}
  {2004})}\BibitemShut {NoStop}%
\bibitem [{\citenamefont {Bruderer}\ \emph {et~al.}(2007)\citenamefont
  {Bruderer}, \citenamefont {Klein}, \citenamefont {Clark},\ and\ \citenamefont
  {Jaksch}}]{bruderer2007polaron}%
  \BibitemOpen
  \bibfield  {author} {\bibinfo {author} {\bibfnamefont {M.}~\bibnamefont
  {Bruderer}}, \bibinfo {author} {\bibfnamefont {A.}~\bibnamefont {Klein}},
  \bibinfo {author} {\bibfnamefont {S.~R.}\ \bibnamefont {Clark}}, \ and\
  \bibinfo {author} {\bibfnamefont {D.}~\bibnamefont {Jaksch}},\ }\bibfield
  {title} {\enquote {\bibinfo {title} {Polaron physics in optical lattices},}\
  }\href@noop {} {\bibfield  {journal} {\bibinfo  {journal} {Physical Review
  A}\ }\textbf {\bibinfo {volume} {76}},\ \bibinfo {pages} {011605} (\bibinfo
  {year} {2007})}\BibitemShut {NoStop}%
\bibitem [{\citenamefont {Fomin}\ \emph {et~al.}(1998)\citenamefont {Fomin},
  \citenamefont {Gladilin}, \citenamefont {Devreese}, \citenamefont
  {Pokatilov}, \citenamefont {Balaban},\ and\ \citenamefont
  {Klimin}}]{fomin1998photoluminescence}%
  \BibitemOpen
  \bibfield  {author} {\bibinfo {author} {\bibfnamefont {V.}~\bibnamefont
  {Fomin}}, \bibinfo {author} {\bibfnamefont {V.}~\bibnamefont {Gladilin}},
  \bibinfo {author} {\bibfnamefont {J.}~\bibnamefont {Devreese}}, \bibinfo
  {author} {\bibfnamefont {E.}~\bibnamefont {Pokatilov}}, \bibinfo {author}
  {\bibfnamefont {S.}~\bibnamefont {Balaban}}, \ and\ \bibinfo {author}
  {\bibfnamefont {S.}~\bibnamefont {Klimin}},\ }\bibfield  {title} {\enquote
  {\bibinfo {title} {Photoluminescence of spherical quantum dots},}\
  }\href@noop {} {\bibfield  {journal} {\bibinfo  {journal} {Physical Review
  B}\ }\textbf {\bibinfo {volume} {57}},\ \bibinfo {pages} {2415} (\bibinfo
  {year} {1998})}\BibitemShut {NoStop}%
\bibitem [{\citenamefont {Klimin}\ \emph {et~al.}(2004)\citenamefont {Klimin},
  \citenamefont {Fomin}, \citenamefont {Brosens},\ and\ \citenamefont
  {Devreese}}]{klimin2004ground}%
  \BibitemOpen
  \bibfield  {author} {\bibinfo {author} {\bibfnamefont {S.}~\bibnamefont
  {Klimin}}, \bibinfo {author} {\bibfnamefont {V.}~\bibnamefont {Fomin}},
  \bibinfo {author} {\bibfnamefont {F.}~\bibnamefont {Brosens}}, \ and\
  \bibinfo {author} {\bibfnamefont {J.}~\bibnamefont {Devreese}},\ }\bibfield
  {title} {\enquote {\bibinfo {title} {Ground state and optical conductivity of
  interacting polarons in a quantum dot},}\ }\href@noop {} {\bibfield
  {journal} {\bibinfo  {journal} {Physical Review B}\ }\textbf {\bibinfo
  {volume} {69}},\ \bibinfo {pages} {235324} (\bibinfo {year}
  {2004})}\BibitemShut {NoStop}%
\bibitem [{\citenamefont {Alexandrov}\ and\ \citenamefont
  {Bratkovsky}(2003)}]{alexandrov2003memory}%
  \BibitemOpen
  \bibfield  {author} {\bibinfo {author} {\bibfnamefont {A.}~\bibnamefont
  {Alexandrov}}\ and\ \bibinfo {author} {\bibfnamefont {A.}~\bibnamefont
  {Bratkovsky}},\ }\bibfield  {title} {\enquote {\bibinfo {title} {Memory
  effect in a molecular quantum dot with strong electron-vibron interaction},}\
  }\href@noop {} {\bibfield  {journal} {\bibinfo  {journal} {Physical Review
  B}\ }\textbf {\bibinfo {volume} {67}},\ \bibinfo {pages} {235312} (\bibinfo
  {year} {2003})}\BibitemShut {NoStop}%
\bibitem [{\citenamefont {Alexandrov}\ and\ \citenamefont
  {Bratkovsky}(2009)}]{alexandrov2009polaronic}%
  \BibitemOpen
  \bibfield  {author} {\bibinfo {author} {\bibfnamefont {A.}~\bibnamefont
  {Alexandrov}}\ and\ \bibinfo {author} {\bibfnamefont {A.}~\bibnamefont
  {Bratkovsky}},\ }\bibfield  {title} {\enquote {\bibinfo {title} {Polaronic
  memory resistors strongly coupled to electrodes},}\ }\href@noop {} {\bibfield
   {journal} {\bibinfo  {journal} {Physical Review B}\ }\textbf {\bibinfo
  {volume} {80}},\ \bibinfo {pages} {115321} (\bibinfo {year}
  {2009})}\BibitemShut {NoStop}%
\bibitem [{\citenamefont {J{\o}rgensen}\ \emph {et~al.}(2016)\citenamefont
  {J{\o}rgensen}, \citenamefont {Wacker}, \citenamefont {Skalmstang},
  \citenamefont {Parish}, \citenamefont {Levinsen}, \citenamefont
  {Christensen}, \citenamefont {Bruun},\ and\ \citenamefont
  {Arlt}}]{jorgensen2016observation}%
  \BibitemOpen
  \bibfield  {author} {\bibinfo {author} {\bibfnamefont {N.~B.}\ \bibnamefont
  {J{\o}rgensen}}, \bibinfo {author} {\bibfnamefont {L.}~\bibnamefont
  {Wacker}}, \bibinfo {author} {\bibfnamefont {K.~T.}\ \bibnamefont
  {Skalmstang}}, \bibinfo {author} {\bibfnamefont {M.~M.}\ \bibnamefont
  {Parish}}, \bibinfo {author} {\bibfnamefont {J.}~\bibnamefont {Levinsen}},
  \bibinfo {author} {\bibfnamefont {R.~S.}\ \bibnamefont {Christensen}},
  \bibinfo {author} {\bibfnamefont {G.~M.}\ \bibnamefont {Bruun}}, \ and\
  \bibinfo {author} {\bibfnamefont {J.~J.}\ \bibnamefont {Arlt}},\ }\bibfield
  {title} {\enquote {\bibinfo {title} {Observation of attractive and repulsive
  polarons in a bose-einstein condensate},}\ }\href@noop {} {\bibfield
  {journal} {\bibinfo  {journal} {Physical review letters}\ }\textbf {\bibinfo
  {volume} {117}},\ \bibinfo {pages} {055302} (\bibinfo {year}
  {2016})}\BibitemShut {NoStop}%
\bibitem [{\citenamefont {Hu}\ \emph {et~al.}(2016)\citenamefont {Hu},
  \citenamefont {Van~de Graaff}, \citenamefont {Kedar}, \citenamefont {Corson},
  \citenamefont {Cornell},\ and\ \citenamefont {Jin}}]{hu2016bose}%
  \BibitemOpen
  \bibfield  {author} {\bibinfo {author} {\bibfnamefont {M.-G.}\ \bibnamefont
  {Hu}}, \bibinfo {author} {\bibfnamefont {M.~J.}\ \bibnamefont {Van~de
  Graaff}}, \bibinfo {author} {\bibfnamefont {D.}~\bibnamefont {Kedar}},
  \bibinfo {author} {\bibfnamefont {J.~P.}\ \bibnamefont {Corson}}, \bibinfo
  {author} {\bibfnamefont {E.~A.}\ \bibnamefont {Cornell}}, \ and\ \bibinfo
  {author} {\bibfnamefont {D.~S.}\ \bibnamefont {Jin}},\ }\bibfield  {title}
  {\enquote {\bibinfo {title} {Bose polarons in the strongly interacting
  regime},}\ }\href@noop {} {\bibfield  {journal} {\bibinfo  {journal}
  {Physical review letters}\ }\textbf {\bibinfo {volume} {117}},\ \bibinfo
  {pages} {055301} (\bibinfo {year} {2016})}\BibitemShut {NoStop}%
\bibitem [{\citenamefont {Casteels}, \citenamefont {Tempere},\ and\
  \citenamefont {Devreese}(2011)}]{casteels2011many}%
  \BibitemOpen
  \bibfield  {author} {\bibinfo {author} {\bibfnamefont {W.}~\bibnamefont
  {Casteels}}, \bibinfo {author} {\bibfnamefont {J.}~\bibnamefont {Tempere}}, \
  and\ \bibinfo {author} {\bibfnamefont {J.}~\bibnamefont {Devreese}},\
  }\bibfield  {title} {\enquote {\bibinfo {title} {Many-polaron description of
  impurities in a bose-einstein condensate in the weak-coupling regime},}\
  }\href@noop {} {\bibfield  {journal} {\bibinfo  {journal} {Physical Review
  A}\ }\textbf {\bibinfo {volume} {84}},\ \bibinfo {pages} {063612} (\bibinfo
  {year} {2011})}\BibitemShut {NoStop}%
\bibitem [{\citenamefont {Roberts}\ and\ \citenamefont
  {Rica}(2009)}]{roberts2009impurity}%
  \BibitemOpen
  \bibfield  {author} {\bibinfo {author} {\bibfnamefont {D.~C.}\ \bibnamefont
  {Roberts}}\ and\ \bibinfo {author} {\bibfnamefont {S.}~\bibnamefont {Rica}},\
  }\bibfield  {title} {\enquote {\bibinfo {title} {Impurity crystal in a
  bose-einstein condensate},}\ }\href@noop {} {\bibfield  {journal} {\bibinfo
  {journal} {Physical review letters}\ }\textbf {\bibinfo {volume} {102}},\
  \bibinfo {pages} {025301} (\bibinfo {year} {2009})}\BibitemShut {NoStop}%
\bibitem [{\citenamefont {Santamore}\ and\ \citenamefont
  {Timmermans}(2011)}]{santamore2011multi}%
  \BibitemOpen
  \bibfield  {author} {\bibinfo {author} {\bibfnamefont {D.}~\bibnamefont
  {Santamore}}\ and\ \bibinfo {author} {\bibfnamefont {E.}~\bibnamefont
  {Timmermans}},\ }\bibfield  {title} {\enquote {\bibinfo {title}
  {Multi-impurity polarons in a dilute bose--einstein condensate},}\
  }\href@noop {} {\bibfield  {journal} {\bibinfo  {journal} {New Journal of
  Physics}\ }\textbf {\bibinfo {volume} {13}},\ \bibinfo {pages} {103029}
  (\bibinfo {year} {2011})}\BibitemShut {NoStop}%
\bibitem [{\citenamefont {Blinova}, \citenamefont {Boshier},\ and\
  \citenamefont {Timmermans}(2013)}]{blinova2013two}%
  \BibitemOpen
  \bibfield  {author} {\bibinfo {author} {\bibfnamefont {A.}~\bibnamefont
  {Blinova}}, \bibinfo {author} {\bibfnamefont {M.}~\bibnamefont {Boshier}}, \
  and\ \bibinfo {author} {\bibfnamefont {E.}~\bibnamefont {Timmermans}},\
  }\bibfield  {title} {\enquote {\bibinfo {title} {Two polaron flavors of the
  bose-einstein condensate impurity},}\ }\href@noop {} {\bibfield  {journal}
  {\bibinfo  {journal} {Physical Review A}\ }\textbf {\bibinfo {volume} {88}},\
  \bibinfo {pages} {053610} (\bibinfo {year} {2013})}\BibitemShut {NoStop}%
\bibitem [{\citenamefont {Casteels}, \citenamefont {Tempere},\ and\
  \citenamefont {Devreese}(2013)}]{casteels2013bipolarons}%
  \BibitemOpen
  \bibfield  {author} {\bibinfo {author} {\bibfnamefont {W.}~\bibnamefont
  {Casteels}}, \bibinfo {author} {\bibfnamefont {J.}~\bibnamefont {Tempere}}, \
  and\ \bibinfo {author} {\bibfnamefont {J.}~\bibnamefont {Devreese}},\
  }\bibfield  {title} {\enquote {\bibinfo {title} {Bipolarons and multipolarons
  consisting of impurity atoms in a bose-einstein condensate},}\ }\href@noop {}
  {\bibfield  {journal} {\bibinfo  {journal} {Physical Review A}\ }\textbf
  {\bibinfo {volume} {88}},\ \bibinfo {pages} {013613} (\bibinfo {year}
  {2013})}\BibitemShut {NoStop}%
\bibitem [{\citenamefont {Utesov}, \citenamefont {Baglay},\ and\ \citenamefont
  {Andreev}(2018)}]{utesov2018effective}%
  \BibitemOpen
  \bibfield  {author} {\bibinfo {author} {\bibfnamefont {O.}~\bibnamefont
  {Utesov}}, \bibinfo {author} {\bibfnamefont {M.}~\bibnamefont {Baglay}}, \
  and\ \bibinfo {author} {\bibfnamefont {S.}~\bibnamefont {Andreev}},\
  }\bibfield  {title} {\enquote {\bibinfo {title} {Effective interactions in a
  quantum bose-bose mixture},}\ }\href@noop {} {\bibfield  {journal} {\bibinfo
  {journal} {Physical Review A}\ }\textbf {\bibinfo {volume} {97}},\ \bibinfo
  {pages} {053617} (\bibinfo {year} {2018})}\BibitemShut {NoStop}%
\bibitem [{\citenamefont {Camacho-Guardian}\ \emph {et~al.}(2018)\citenamefont
  {Camacho-Guardian}, \citenamefont {Ardila}, \citenamefont {Pohl},\ and\
  \citenamefont {Bruun}}]{camacho2018bipolarons}%
  \BibitemOpen
  \bibfield  {author} {\bibinfo {author} {\bibfnamefont {A.}~\bibnamefont
  {Camacho-Guardian}}, \bibinfo {author} {\bibfnamefont {L.~A.~P.}\
  \bibnamefont {Ardila}}, \bibinfo {author} {\bibfnamefont {T.}~\bibnamefont
  {Pohl}}, \ and\ \bibinfo {author} {\bibfnamefont {G.~M.}\ \bibnamefont
  {Bruun}},\ }\bibfield  {title} {\enquote {\bibinfo {title} {Bipolarons in a
  bose-einstein condensate},}\ }\href@noop {} {\bibfield  {journal} {\bibinfo
  {journal} {arXiv preprint arXiv:1804.00402}\ } (\bibinfo {year}
  {2018})}\BibitemShut {NoStop}%
\bibitem [{\citenamefont {Bijlsma}, \citenamefont {Heringa},\ and\
  \citenamefont {Stoof}(2000)}]{bijlsma2000phonon}%
  \BibitemOpen
  \bibfield  {author} {\bibinfo {author} {\bibfnamefont {M.}~\bibnamefont
  {Bijlsma}}, \bibinfo {author} {\bibfnamefont {B.}~\bibnamefont {Heringa}}, \
  and\ \bibinfo {author} {\bibfnamefont {H.}~\bibnamefont {Stoof}},\ }\bibfield
   {title} {\enquote {\bibinfo {title} {Phonon exchange in dilute fermi-bose
  mixtures: Tailoring the fermi-fermi interaction},}\ }\href@noop {} {\bibfield
   {journal} {\bibinfo  {journal} {Physical Review A}\ }\textbf {\bibinfo
  {volume} {61}},\ \bibinfo {pages} {053601} (\bibinfo {year}
  {2000})}\BibitemShut {NoStop}%
\bibitem [{\citenamefont {Ruderman}\ and\ \citenamefont
  {Kittel}(1954)}]{ruderman1954indirect}%
  \BibitemOpen
  \bibfield  {author} {\bibinfo {author} {\bibfnamefont {M.~A.}\ \bibnamefont
  {Ruderman}}\ and\ \bibinfo {author} {\bibfnamefont {C.}~\bibnamefont
  {Kittel}},\ }\bibfield  {title} {\enquote {\bibinfo {title} {Indirect
  exchange coupling of nuclear magnetic moments by conduction electrons},}\
  }\href@noop {} {\bibfield  {journal} {\bibinfo  {journal} {Physical Review}\
  }\textbf {\bibinfo {volume} {96}},\ \bibinfo {pages} {99} (\bibinfo {year}
  {1954})}\BibitemShut {NoStop}%
\bibitem [{\citenamefont {Zhou}\ \emph {et~al.}(2010)\citenamefont {Zhou},
  \citenamefont {Wiebe}, \citenamefont {Lounis}, \citenamefont {Vedmedenko},
  \citenamefont {Meier}, \citenamefont {Bl{\"u}gel}, \citenamefont
  {Dederichs},\ and\ \citenamefont {Wiesendanger}}]{zhou2010strength}%
  \BibitemOpen
  \bibfield  {author} {\bibinfo {author} {\bibfnamefont {L.}~\bibnamefont
  {Zhou}}, \bibinfo {author} {\bibfnamefont {J.}~\bibnamefont {Wiebe}},
  \bibinfo {author} {\bibfnamefont {S.}~\bibnamefont {Lounis}}, \bibinfo
  {author} {\bibfnamefont {E.}~\bibnamefont {Vedmedenko}}, \bibinfo {author}
  {\bibfnamefont {F.}~\bibnamefont {Meier}}, \bibinfo {author} {\bibfnamefont
  {S.}~\bibnamefont {Bl{\"u}gel}}, \bibinfo {author} {\bibfnamefont {P.~H.}\
  \bibnamefont {Dederichs}}, \ and\ \bibinfo {author} {\bibfnamefont
  {R.}~\bibnamefont {Wiesendanger}},\ }\bibfield  {title} {\enquote {\bibinfo
  {title} {Strength and directionality of surface
  ruderman--kittel--kasuya--yosida interaction mapped on the atomic scale},}\
  }\href@noop {} {\bibfield  {journal} {\bibinfo  {journal} {Nature Physics}\
  }\textbf {\bibinfo {volume} {6}},\ \bibinfo {pages} {187} (\bibinfo {year}
  {2010})}\BibitemShut {NoStop}%
\bibitem [{\citenamefont {Hewson}(1997)}]{hewson1997kondo}%
  \BibitemOpen
  \bibfield  {author} {\bibinfo {author} {\bibfnamefont {A.~C.}\ \bibnamefont
  {Hewson}},\ }\href@noop {} {\emph {\bibinfo {title} {The Kondo problem to
  heavy fermions}}},\ Vol.~\bibinfo {volume} {2}\ (\bibinfo  {publisher}
  {Cambridge university press},\ \bibinfo {year} {1997})\BibitemShut {NoStop}%
\bibitem [{\citenamefont {Schmidt}\ and\ \citenamefont
  {Lemeshko}(2015)}]{schmidt2015rotation}%
  \BibitemOpen
  \bibfield  {author} {\bibinfo {author} {\bibfnamefont {R.}~\bibnamefont
  {Schmidt}}\ and\ \bibinfo {author} {\bibfnamefont {M.}~\bibnamefont
  {Lemeshko}},\ }\bibfield  {title} {\enquote {\bibinfo {title} {Rotation of
  quantum impurities in the presence of a many-body environment},}\ }\href@noop
  {} {\bibfield  {journal} {\bibinfo  {journal} {Physical review letters}\
  }\textbf {\bibinfo {volume} {114}},\ \bibinfo {pages} {203001} (\bibinfo
  {year} {2015})}\BibitemShut {NoStop}%
\bibitem [{\citenamefont {Schmidt}\ and\ \citenamefont
  {Lemeshko}(2016)}]{schmidt2016deformation}%
  \BibitemOpen
  \bibfield  {author} {\bibinfo {author} {\bibfnamefont {R.}~\bibnamefont
  {Schmidt}}\ and\ \bibinfo {author} {\bibfnamefont {M.}~\bibnamefont
  {Lemeshko}},\ }\bibfield  {title} {\enquote {\bibinfo {title} {Deformation of
  a quantum many-particle system by a rotating impurity},}\ }\href@noop {}
  {\bibfield  {journal} {\bibinfo  {journal} {Physical Review X}\ }\textbf
  {\bibinfo {volume} {6}},\ \bibinfo {pages} {011012} (\bibinfo {year}
  {2016})}\BibitemShut {NoStop}%
\bibitem [{\citenamefont {Lemeshko}\ and\ \citenamefont
  {Schmidt}(2017)}]{lemeshko2017molecular}%
  \BibitemOpen
  \bibfield  {author} {\bibinfo {author} {\bibfnamefont {M.}~\bibnamefont
  {Lemeshko}}\ and\ \bibinfo {author} {\bibfnamefont {R.}~\bibnamefont
  {Schmidt}},\ }\bibfield  {title} {\enquote {\bibinfo {title} {Molecular
  impurities interacting with a many-particle environment: from ultracold gases
  to helium nanodroplets},}\ }\href@noop {} {\bibfield  {journal} {\bibinfo
  {journal} {Low Energy and Low Temperature Molecular Scattering}\ } (\bibinfo
  {year} {2017})}\BibitemShut {NoStop}%
\bibitem [{\citenamefont {Bighin}, \citenamefont {Tscherbul},\ and\
  \citenamefont {Lemeshko}(2018)}]{bighin2018diagrammaticprl}%
  \BibitemOpen
  \bibfield  {author} {\bibinfo {author} {\bibfnamefont {G.}~\bibnamefont
  {Bighin}}, \bibinfo {author} {\bibfnamefont {T.}~\bibnamefont {Tscherbul}}, \
  and\ \bibinfo {author} {\bibfnamefont {M.}~\bibnamefont {Lemeshko}},\
  }\bibfield  {title} {\enquote {\bibinfo {title} {Diagrammatic monte carlo
  approach to rotating molecular impurities},}\ }\href@noop {} {\bibfield
  {journal} {\bibinfo  {journal} {Physical Review Letters}\ }\textbf {\bibinfo
  {volume} {121}} (\bibinfo {year} {2018})}\BibitemShut {NoStop}%
\bibitem [{\citenamefont {Lemeshko}(2017)}]{lemeshko2017quasiparticle}%
  \BibitemOpen
  \bibfield  {author} {\bibinfo {author} {\bibfnamefont {M.}~\bibnamefont
  {Lemeshko}},\ }\bibfield  {title} {\enquote {\bibinfo {title} {Quasiparticle
  approach to molecules interacting with quantum solvents},}\ }\href@noop {}
  {\bibfield  {journal} {\bibinfo  {journal} {Physical review letters}\
  }\textbf {\bibinfo {volume} {118}},\ \bibinfo {pages} {095301} (\bibinfo
  {year} {2017})}\BibitemShut {NoStop}%
\bibitem [{\citenamefont {Shepperson}\ \emph
  {et~al.}(2017{\natexlab{a}})\citenamefont {Shepperson}, \citenamefont
  {S{\o}ndergaard}, \citenamefont {Christiansen}, \citenamefont {Kaczmarczyk},
  \citenamefont {Zillich}, \citenamefont {Lemeshko},\ and\ \citenamefont
  {Stapelfeldt}}]{shepperson2017laser}%
  \BibitemOpen
  \bibfield  {author} {\bibinfo {author} {\bibfnamefont {B.}~\bibnamefont
  {Shepperson}}, \bibinfo {author} {\bibfnamefont {A.~A.}\ \bibnamefont
  {S{\o}ndergaard}}, \bibinfo {author} {\bibfnamefont {L.}~\bibnamefont
  {Christiansen}}, \bibinfo {author} {\bibfnamefont {J.}~\bibnamefont
  {Kaczmarczyk}}, \bibinfo {author} {\bibfnamefont {R.~E.}\ \bibnamefont
  {Zillich}}, \bibinfo {author} {\bibfnamefont {M.}~\bibnamefont {Lemeshko}}, \
  and\ \bibinfo {author} {\bibfnamefont {H.}~\bibnamefont {Stapelfeldt}},\
  }\bibfield  {title} {\enquote {\bibinfo {title} {Laser-induced rotation of
  iodine molecules in helium nanodroplets: revivals and breaking free},}\
  }\href@noop {} {\bibfield  {journal} {\bibinfo  {journal} {Physical review
  letters}\ }\textbf {\bibinfo {volume} {118}},\ \bibinfo {pages} {203203}
  (\bibinfo {year} {2017}{\natexlab{a}})}\BibitemShut {NoStop}%
\bibitem [{\citenamefont {Cherepanov}\ and\ \citenamefont
  {Lemeshko}(2017)}]{cherepanov2017fingerprints}%
  \BibitemOpen
  \bibfield  {author} {\bibinfo {author} {\bibfnamefont {I.~N.}\ \bibnamefont
  {Cherepanov}}\ and\ \bibinfo {author} {\bibfnamefont {M.}~\bibnamefont
  {Lemeshko}},\ }\bibfield  {title} {\enquote {\bibinfo {title} {Fingerprints
  of angulon instabilities in the spectra of matrix-isolated molecules},}\
  }\href@noop {} {\bibfield  {journal} {\bibinfo  {journal} {Physical Review
  Materials}\ }\textbf {\bibinfo {volume} {1}},\ \bibinfo {pages} {035602}
  (\bibinfo {year} {2017})}\BibitemShut {NoStop}%
\bibitem [{\citenamefont {Shepperson}\ \emph
  {et~al.}(2017{\natexlab{b}})\citenamefont {Shepperson}, \citenamefont
  {Chatterley}, \citenamefont {S{\o}ndergaard}, \citenamefont {Christiansen},
  \citenamefont {Lemeshko},\ and\ \citenamefont
  {Stapelfeldt}}]{shepperson2017strongly}%
  \BibitemOpen
  \bibfield  {author} {\bibinfo {author} {\bibfnamefont {B.}~\bibnamefont
  {Shepperson}}, \bibinfo {author} {\bibfnamefont {A.~S.}\ \bibnamefont
  {Chatterley}}, \bibinfo {author} {\bibfnamefont {A.~A.}\ \bibnamefont
  {S{\o}ndergaard}}, \bibinfo {author} {\bibfnamefont {L.}~\bibnamefont
  {Christiansen}}, \bibinfo {author} {\bibfnamefont {M.}~\bibnamefont
  {Lemeshko}}, \ and\ \bibinfo {author} {\bibfnamefont {H.}~\bibnamefont
  {Stapelfeldt}},\ }\bibfield  {title} {\enquote {\bibinfo {title} {Strongly
  aligned molecules inside helium droplets in the near-adiabatic regime},}\
  }\href@noop {} {\bibfield  {journal} {\bibinfo  {journal} {The Journal of
  Chemical Physics}\ }\textbf {\bibinfo {volume} {147}},\ \bibinfo {pages}
  {013946} (\bibinfo {year} {2017}{\natexlab{b}})}\BibitemShut {NoStop}%
\bibitem [{\citenamefont {Cherepanov}\ \emph {et~al.}(2019)\citenamefont
  {Cherepanov}, \citenamefont {Bighin}, \citenamefont {Christiansen},
  \citenamefont {J{\o}rgensen}, \citenamefont {Schmidt}, \citenamefont
  {Stapelfeldt},\ and\ \citenamefont {Lemeshko}}]{cherepanov2019far}%
  \BibitemOpen
  \bibfield  {author} {\bibinfo {author} {\bibfnamefont {I.~N.}\ \bibnamefont
  {Cherepanov}}, \bibinfo {author} {\bibfnamefont {G.}~\bibnamefont {Bighin}},
  \bibinfo {author} {\bibfnamefont {L.}~\bibnamefont {Christiansen}}, \bibinfo
  {author} {\bibfnamefont {A.~V.}\ \bibnamefont {J{\o}rgensen}}, \bibinfo
  {author} {\bibfnamefont {R.}~\bibnamefont {Schmidt}}, \bibinfo {author}
  {\bibfnamefont {H.}~\bibnamefont {Stapelfeldt}}, \ and\ \bibinfo {author}
  {\bibfnamefont {M.}~\bibnamefont {Lemeshko}},\ }\bibfield  {title} {\enquote
  {\bibinfo {title} {Far-from-equilibrium dynamics of angular momentum in a
  quantum many-particle system},}\ }\href@noop {} {\bibfield  {journal}
  {\bibinfo  {journal} {arXiv preprint arXiv:1906.12238}\ } (\bibinfo {year}
  {2019})}\BibitemShut {NoStop}%
\bibitem [{\citenamefont {Yakaboylu}, \citenamefont {Deuchert},\ and\
  \citenamefont {Lemeshko}(2017)}]{yakaboylu2017emergence}%
  \BibitemOpen
  \bibfield  {author} {\bibinfo {author} {\bibfnamefont {E.}~\bibnamefont
  {Yakaboylu}}, \bibinfo {author} {\bibfnamefont {A.}~\bibnamefont {Deuchert}},
  \ and\ \bibinfo {author} {\bibfnamefont {M.}~\bibnamefont {Lemeshko}},\
  }\bibfield  {title} {\enquote {\bibinfo {title} {Emergence of non-abelian
  magnetic monopoles in a quantum impurity problem},}\ }\href@noop {}
  {\bibfield  {journal} {\bibinfo  {journal} {Physical review letters}\
  }\textbf {\bibinfo {volume} {119}},\ \bibinfo {pages} {235301} (\bibinfo
  {year} {2017})}\BibitemShut {NoStop}%
\bibitem [{\citenamefont {Yakaboylu}\ and\ \citenamefont
  {Lemeshko}(2017)}]{yakaboylu2017anomalous}%
  \BibitemOpen
  \bibfield  {author} {\bibinfo {author} {\bibfnamefont {E.}~\bibnamefont
  {Yakaboylu}}\ and\ \bibinfo {author} {\bibfnamefont {M.}~\bibnamefont
  {Lemeshko}},\ }\bibfield  {title} {\enquote {\bibinfo {title} {Anomalous
  screening of quantum impurities by a neutral environment},}\ }\href@noop {}
  {\bibfield  {journal} {\bibinfo  {journal} {Physical review letters}\
  }\textbf {\bibinfo {volume} {118}},\ \bibinfo {pages} {085302} (\bibinfo
  {year} {2017})}\BibitemShut {NoStop}%
\bibitem [{\citenamefont {Toennies}\ and\ \citenamefont
  {Vilesov}(2004)}]{Toennies2004}%
  \BibitemOpen
  \bibfield  {author} {\bibinfo {author} {\bibfnamefont {J.~P.}\ \bibnamefont
  {Toennies}}\ and\ \bibinfo {author} {\bibfnamefont {F.~A.}\ \bibnamefont
  {Vilesov}},\ }\bibfield  {title} {\enquote {\bibinfo {title} {Superfluid
  helium droplets: A uniquely cold nanomatrix for molecules and molecular
  complexes},}\ }\href@noop {} {\bibfield  {journal} {\bibinfo  {journal}
  {Angewandte Chemie}\ }\textbf {\bibinfo {volume} {43}},\ \bibinfo {pages}
  {2622} (\bibinfo {year} {2004})}\BibitemShut {NoStop}%
\bibitem [{\citenamefont {Pickering}\ \emph {et~al.}(2018)\citenamefont
  {Pickering}, \citenamefont {Shepperson}, \citenamefont {H\"ubschmann},
  \citenamefont {Thorning},\ and\ \citenamefont
  {Stapelfeldt}}]{PhysRevLett.120.113202}%
  \BibitemOpen
  \bibfield  {author} {\bibinfo {author} {\bibfnamefont {J.~D.}\ \bibnamefont
  {Pickering}}, \bibinfo {author} {\bibfnamefont {B.}~\bibnamefont
  {Shepperson}}, \bibinfo {author} {\bibfnamefont {B.~A.~K.}\ \bibnamefont
  {H\"ubschmann}}, \bibinfo {author} {\bibfnamefont {F.}~\bibnamefont
  {Thorning}}, \ and\ \bibinfo {author} {\bibfnamefont {H.}~\bibnamefont
  {Stapelfeldt}},\ }\bibfield  {title} {\enquote {\bibinfo {title} {Alignment
  and imaging of the ${\mathrm{cs}}_{2}$ dimer inside helium nanodroplets},}\
  }\href {\doibase 10.1103/PhysRevLett.120.113202} {\bibfield  {journal}
  {\bibinfo  {journal} {Phys. Rev. Lett.}\ }\textbf {\bibinfo {volume} {120}},\
  \bibinfo {pages} {113202} (\bibinfo {year} {2018})}\BibitemShut {NoStop}%
\bibitem [{\citenamefont {Fischer}\ \emph {et~al.}(0)\citenamefont {Fischer},
  \citenamefont {Schlaghaufer}, \citenamefont {Lottner}, \citenamefont
  {Slenczka}, \citenamefont {Christiansen}, \citenamefont {Stapelfeldt},
  \citenamefont {Karra}, \citenamefont {Friedrich}, \citenamefont {Mullan},
  \citenamefont {Schütz},\ and\ \citenamefont
  {Usvyat}}]{doi:10.1021/acs.jpca.9b07302}%
  \BibitemOpen
  \bibfield  {author} {\bibinfo {author} {\bibfnamefont {J.}~\bibnamefont
  {Fischer}}, \bibinfo {author} {\bibfnamefont {F.}~\bibnamefont
  {Schlaghaufer}}, \bibinfo {author} {\bibfnamefont {E.-M.}\ \bibnamefont
  {Lottner}}, \bibinfo {author} {\bibfnamefont {A.}~\bibnamefont {Slenczka}},
  \bibinfo {author} {\bibfnamefont {L.}~\bibnamefont {Christiansen}}, \bibinfo
  {author} {\bibfnamefont {H.}~\bibnamefont {Stapelfeldt}}, \bibinfo {author}
  {\bibfnamefont {M.}~\bibnamefont {Karra}}, \bibinfo {author} {\bibfnamefont
  {B.}~\bibnamefont {Friedrich}}, \bibinfo {author} {\bibfnamefont
  {T.}~\bibnamefont {Mullan}}, \bibinfo {author} {\bibfnamefont
  {M.}~\bibnamefont {Schütz}}, \ and\ \bibinfo {author} {\bibfnamefont
  {D.}~\bibnamefont {Usvyat}},\ }\bibfield  {title} {\enquote {\bibinfo {title}
  {Heterogeneous clusters of phthalocyanine and water prepared and probed in
  superfluid helium nanodroplets},}\ }\href {\doibase 10.1021/acs.jpca.9b07302}
  {\bibfield  {journal} {\bibinfo  {journal} {The Journal of Physical Chemistry
  A}\ }\textbf {\bibinfo {volume} {0}},\ \bibinfo {pages} {null} (\bibinfo
  {year} {0})}\BibitemShut {NoStop}%
\bibitem [{\citenamefont {Sadoon}\ \emph {et~al.}(2016)\citenamefont {Sadoon},
  \citenamefont {Sarma}, \citenamefont {Cunningham}, \citenamefont {Tandy},
  \citenamefont {Hanson-Heine}, \citenamefont {Besley}, \citenamefont {Yang},\
  and\ \citenamefont {Ellis}}]{doi:10.1021/acs.jpca.6b06227}%
  \BibitemOpen
  \bibfield  {author} {\bibinfo {author} {\bibfnamefont {A.~M.}\ \bibnamefont
  {Sadoon}}, \bibinfo {author} {\bibfnamefont {G.}~\bibnamefont {Sarma}},
  \bibinfo {author} {\bibfnamefont {E.~M.}\ \bibnamefont {Cunningham}},
  \bibinfo {author} {\bibfnamefont {J.}~\bibnamefont {Tandy}}, \bibinfo
  {author} {\bibfnamefont {M.~W.~D.}\ \bibnamefont {Hanson-Heine}}, \bibinfo
  {author} {\bibfnamefont {N.~A.}\ \bibnamefont {Besley}}, \bibinfo {author}
  {\bibfnamefont {S.}~\bibnamefont {Yang}}, \ and\ \bibinfo {author}
  {\bibfnamefont {A.~M.}\ \bibnamefont {Ellis}},\ }\bibfield  {title} {\enquote
  {\bibinfo {title} {Infrared spectroscopy of nacl(ch3oh)n complexes in helium
  nanodroplets},}\ }\href {\doibase 10.1021/acs.jpca.6b06227} {\bibfield
  {journal} {\bibinfo  {journal} {The Journal of Physical Chemistry A}\
  }\textbf {\bibinfo {volume} {120}},\ \bibinfo {pages} {8085--8092} (\bibinfo
  {year} {2016})}\BibitemShut {NoStop}%
\bibitem [{\citenamefont {Pekar}(1946)}]{pekar1946local}%
  \BibitemOpen
  \bibfield  {author} {\bibinfo {author} {\bibfnamefont {S.}~\bibnamefont
  {Pekar}},\ }\bibfield  {title} {\enquote {\bibinfo {title} {Local quantum
  states of electrons in an ideal ion crystal},}\ }\href@noop {} {\bibfield
  {journal} {\bibinfo  {journal} {Zhurnal Eksperimentalnoi I Teoreticheskoi
  Fiziki}\ }\textbf {\bibinfo {volume} {16}},\ \bibinfo {pages} {341--348}
  (\bibinfo {year} {1946})}\BibitemShut {NoStop}%
\bibitem [{\citenamefont {Stone}(2013{\natexlab{a}})}]{StoneBook13}%
  \BibitemOpen
  \bibfield  {author} {\bibinfo {author} {\bibfnamefont {A.}~\bibnamefont
  {Stone}},\ }\href@noop {} {\emph {\bibinfo {title} {The Theory of
  Intermolecular Forces}}}\ (\bibinfo  {publisher} {Oxford University Press},\
  \bibinfo {year} {2013})\BibitemShut {NoStop}%
\bibitem [{\citenamefont {Yakaboylu}\ \emph {et~al.}(2018)\citenamefont
  {Yakaboylu}, \citenamefont {Midya}, \citenamefont {Deuchert}, \citenamefont
  {Leopold},\ and\ \citenamefont {Lemeshko}}]{yakaboylu2018theory}%
  \BibitemOpen
  \bibfield  {author} {\bibinfo {author} {\bibfnamefont {E.}~\bibnamefont
  {Yakaboylu}}, \bibinfo {author} {\bibfnamefont {B.}~\bibnamefont {Midya}},
  \bibinfo {author} {\bibfnamefont {A.}~\bibnamefont {Deuchert}}, \bibinfo
  {author} {\bibfnamefont {N.}~\bibnamefont {Leopold}}, \ and\ \bibinfo
  {author} {\bibfnamefont {M.}~\bibnamefont {Lemeshko}},\ }\bibfield  {title}
  {\enquote {\bibinfo {title} {Theory of the rotating polaron: Spectrum and
  self-localization},}\ }\href@noop {} {\bibfield  {journal} {\bibinfo
  {journal} {Physical Review B}\ }\textbf {\bibinfo {volume} {98}},\ \bibinfo
  {pages} {224506} (\bibinfo {year} {2018})}\BibitemShut {NoStop}%
\bibitem [{\citenamefont {Midya}\ \emph {et~al.}(2016)\citenamefont {Midya},
  \citenamefont {Tomza}, \citenamefont {Schmidt},\ and\ \citenamefont
  {Lemeshko}}]{Midya2016}%
  \BibitemOpen
  \bibfield  {author} {\bibinfo {author} {\bibfnamefont {B.}~\bibnamefont
  {Midya}}, \bibinfo {author} {\bibfnamefont {M.}~\bibnamefont {Tomza}},
  \bibinfo {author} {\bibfnamefont {R.}~\bibnamefont {Schmidt}}, \ and\
  \bibinfo {author} {\bibfnamefont {M.}~\bibnamefont {Lemeshko}},\ }\bibfield
  {title} {\enquote {\bibinfo {title} {Rotation of cold molecular ions inside a
  bose-einstein condensate},}\ }\href@noop {} {\bibfield  {journal} {\bibinfo
  {journal} {Physical Review A}\ }\textbf {\bibinfo {volume} {94}},\ \bibinfo
  {pages} {041601} (\bibinfo {year} {2016})}\BibitemShut {NoStop}%
\bibitem [{\citenamefont {Salje}, \citenamefont {Alexandrov},\ and\
  \citenamefont {Liang}(2005)}]{salje2005polarons}%
  \BibitemOpen
  \bibfield  {author} {\bibinfo {author} {\bibfnamefont {E.~K.}\ \bibnamefont
  {Salje}}, \bibinfo {author} {\bibfnamefont {A.}~\bibnamefont {Alexandrov}}, \
  and\ \bibinfo {author} {\bibfnamefont {W.}~\bibnamefont {Liang}},\
  }\href@noop {} {\emph {\bibinfo {title} {Polarons and bipolarons in high-Tc
  superconductors and related materials}}}\ (\bibinfo  {publisher} {Cambridge
  University Press},\ \bibinfo {year} {2005})\BibitemShut {NoStop}%
\bibitem [{\citenamefont {Stone}(2013{\natexlab{b}})}]{stone2013theory}%
  \BibitemOpen
  \bibfield  {author} {\bibinfo {author} {\bibfnamefont {A.}~\bibnamefont
  {Stone}},\ }\href@noop {} {\emph {\bibinfo {title} {The theory of
  intermolecular forces}}}\ (\bibinfo  {publisher} {oUP oxford},\ \bibinfo
  {year} {2013})\BibitemShut {NoStop}%
\bibitem [{\citenamefont {Donnelly}\ and\ \citenamefont
  {Barenghi}(1998)}]{donnelly1998observed}%
  \BibitemOpen
  \bibfield  {author} {\bibinfo {author} {\bibfnamefont {R.~J.}\ \bibnamefont
  {Donnelly}}\ and\ \bibinfo {author} {\bibfnamefont {C.~F.}\ \bibnamefont
  {Barenghi}},\ }\bibfield  {title} {\enquote {\bibinfo {title} {The observed
  properties of liquid helium at the saturated vapor pressure},}\ }\href@noop
  {} {\bibfield  {journal} {\bibinfo  {journal} {Journal of physical and
  chemical reference data}\ }\textbf {\bibinfo {volume} {27}},\ \bibinfo
  {pages} {1217--1274} (\bibinfo {year} {1998})}\BibitemShut {NoStop}%
\bibitem [{\citenamefont {Donsker}\ and\ \citenamefont
  {Varadhan}(1983)}]{Pekarderivation1}%
  \BibitemOpen
  \bibfield  {author} {\bibinfo {author} {\bibfnamefont {M.}~\bibnamefont
  {Donsker}}\ and\ \bibinfo {author} {\bibfnamefont {S.}~\bibnamefont
  {Varadhan}},\ }\bibfield  {title} {\enquote {\bibinfo {title} {Asymptotics
  for the polaron},}\ }\href@noop {} {\bibfield  {journal} {\bibinfo  {journal}
  {Comm. Pure Appl. Math.}\ }\textbf {\bibinfo {volume} {36}},\ \bibinfo
  {pages} {505–528} (\bibinfo {year} {1983})}\BibitemShut {NoStop}%
\bibitem [{\citenamefont {Lieb}\ and\ \citenamefont
  {Thomas}(1997)}]{Pekarderivation2}%
  \BibitemOpen
  \bibfield  {author} {\bibinfo {author} {\bibfnamefont {E.}~\bibnamefont
  {Lieb}}\ and\ \bibinfo {author} {\bibfnamefont {L.~E.}\ \bibnamefont
  {Thomas}},\ }\bibfield  {title} {\enquote {\bibinfo {title} {Exact ground
  state energy of the strong-coupling polaron},}\ }\href@noop {} {\bibfield
  {journal} {\bibinfo  {journal} {Commun. Math. Phys.}\ }\textbf {\bibinfo
  {volume} {183}},\ \bibinfo {pages} {511--519} (\bibinfo {year}
  {1997})}\BibitemShut {NoStop}%
\bibitem [{\citenamefont {Das}\ and\ \citenamefont
  {Chakrabarti}(2005)}]{das2005quantum}%
  \BibitemOpen
  \bibfield  {author} {\bibinfo {author} {\bibfnamefont {A.}~\bibnamefont
  {Das}}\ and\ \bibinfo {author} {\bibfnamefont {B.~K.}\ \bibnamefont
  {Chakrabarti}},\ }\href@noop {} {\emph {\bibinfo {title} {Quantum annealing
  and related optimization methods}}},\ Vol.\ \bibinfo {volume} {679}\
  (\bibinfo  {publisher} {Springer Science \& Business Media},\ \bibinfo {year}
  {2005})\BibitemShut {NoStop}%
\bibitem [{\citenamefont {Abramowitz}\ and\ \citenamefont
  {Stegun}(1965)}]{abramowitz1965handbook}%
  \BibitemOpen
  \bibfield  {author} {\bibinfo {author} {\bibfnamefont {M.}~\bibnamefont
  {Abramowitz}}\ and\ \bibinfo {author} {\bibfnamefont {I.~A.}\ \bibnamefont
  {Stegun}},\ }\href@noop {} {\emph {\bibinfo {title} {Handbook of mathematical
  functions: with formulas, graphs, and mathematical tables}}},\ Vol.~\bibinfo
  {volume} {55}\ (\bibinfo  {publisher} {Courier Corporation},\ \bibinfo {year}
  {1965})\BibitemShut {NoStop}%
\bibitem [{\citenamefont {Reed}\ and\ \citenamefont
  {Simon}(1980)}]{ReedSimon1980}%
  \BibitemOpen
  \bibfield  {author} {\bibinfo {author} {\bibfnamefont {M.}~\bibnamefont
  {Reed}}\ and\ \bibinfo {author} {\bibfnamefont {B.}~\bibnamefont {Simon}},\
  }\href@noop {} {\emph {\bibinfo {title} {Methods of modern mathematical
  physics I, Functional Analysis}}}\ (\bibinfo  {publisher} {Academic Press},\
  \bibinfo {year} {1980})\BibitemShut {NoStop}%
\bibitem [{\citenamefont {Chevy}(2006)}]{chevy2006universal}%
  \BibitemOpen
  \bibfield  {author} {\bibinfo {author} {\bibfnamefont {F.}~\bibnamefont
  {Chevy}},\ }\bibfield  {title} {\enquote {\bibinfo {title} {Universal phase
  diagram of a strongly interacting fermi gas with unbalanced spin
  populations},}\ }\href@noop {} {\bibfield  {journal} {\bibinfo  {journal}
  {Physical Review A}\ }\textbf {\bibinfo {volume} {74}},\ \bibinfo {pages}
  {063628} (\bibinfo {year} {2006})}\BibitemShut {NoStop}%
\bibitem [{\citenamefont {Lan}\ and\ \citenamefont
  {Lobo}(2014)}]{lan2014single}%
  \BibitemOpen
  \bibfield  {author} {\bibinfo {author} {\bibfnamefont {Z.}~\bibnamefont
  {Lan}}\ and\ \bibinfo {author} {\bibfnamefont {C.}~\bibnamefont {Lobo}},\
  }\bibfield  {title} {\enquote {\bibinfo {title} {A single impurity in an
  ideal atomic fermi gas: current understanding and some open problems},}\
  }\href@noop {} {\bibfield  {journal} {\bibinfo  {journal} {arXiv preprint
  arXiv:1404.3220}\ } (\bibinfo {year} {2014})}\BibitemShut {NoStop}%
\bibitem [{\citenamefont {Varshalovich}, \citenamefont {Moskalev},\ and\
  \citenamefont {Khersonskii}(1988)}]{varshalovich1988quantum}%
  \BibitemOpen
  \bibfield  {author} {\bibinfo {author} {\bibfnamefont {D.~A.}\ \bibnamefont
  {Varshalovich}}, \bibinfo {author} {\bibfnamefont {A.~N.}\ \bibnamefont
  {Moskalev}}, \ and\ \bibinfo {author} {\bibfnamefont {V.~K.}\ \bibnamefont
  {Khersonskii}},\ }\href@noop {} {\emph {\bibinfo {title} {Quantum theory of
  angular momentum}}}\ (\bibinfo  {publisher} {World scientific},\ \bibinfo
  {year} {1988})\BibitemShut {NoStop}%
\bibitem [{\citenamefont {Schmidt}, \citenamefont {Sadeghpour},\ and\
  \citenamefont {Demler}(2016)}]{camargo2016creation}%
  \BibitemOpen
  \bibfield  {author} {\bibinfo {author} {\bibfnamefont {R.}~\bibnamefont
  {Schmidt}}, \bibinfo {author} {\bibfnamefont {H.}~\bibnamefont {Sadeghpour}},
  \ and\ \bibinfo {author} {\bibfnamefont {E.}~\bibnamefont {Demler}},\
  }\bibfield  {title} {\enquote {\bibinfo {title} {Mesoscopic rydberg impurity
  in an atomic quantum gas},}\ }\href@noop {} {\bibfield  {journal} {\bibinfo
  {journal} {Physical review letters}\ }\textbf {\bibinfo {volume} {116}},\
  \bibinfo {pages} {105302} (\bibinfo {year} {2016})}\BibitemShut {NoStop}%
\bibitem [{\citenamefont {Camargo}\ \emph {et~al.}(2018)\citenamefont
  {Camargo}, \citenamefont {Schmidt}, \citenamefont {Whalen}, \citenamefont
  {Ding}, \citenamefont {Woehl~Jr}, \citenamefont {Yoshida}, \citenamefont
  {Burgd{\"o}rfer}, \citenamefont {Dunning}, \citenamefont {Sadeghpour},
  \citenamefont {Demler} \emph {et~al.}}]{camargo2018creation}%
  \BibitemOpen
  \bibfield  {author} {\bibinfo {author} {\bibfnamefont {F.}~\bibnamefont
  {Camargo}}, \bibinfo {author} {\bibfnamefont {R.}~\bibnamefont {Schmidt}},
  \bibinfo {author} {\bibfnamefont {J.}~\bibnamefont {Whalen}}, \bibinfo
  {author} {\bibfnamefont {R.}~\bibnamefont {Ding}}, \bibinfo {author}
  {\bibfnamefont {G.}~\bibnamefont {Woehl~Jr}}, \bibinfo {author}
  {\bibfnamefont {S.}~\bibnamefont {Yoshida}}, \bibinfo {author} {\bibfnamefont
  {J.}~\bibnamefont {Burgd{\"o}rfer}}, \bibinfo {author} {\bibfnamefont
  {F.}~\bibnamefont {Dunning}}, \bibinfo {author} {\bibfnamefont
  {H.}~\bibnamefont {Sadeghpour}}, \bibinfo {author} {\bibfnamefont
  {E.}~\bibnamefont {Demler}},  \emph {et~al.},\ }\bibfield  {title} {\enquote
  {\bibinfo {title} {Creation of rydberg polarons in a bose gas},}\ }\href@noop
  {} {\bibfield  {journal} {\bibinfo  {journal} {Physical review letters}\
  }\textbf {\bibinfo {volume} {120}},\ \bibinfo {pages} {083401} (\bibinfo
  {year} {2018})}\BibitemShut {NoStop}%
\bibitem [{\citenamefont {Pushkarov}(1991)}]{pushkarov1991quasiparticle}%
  \BibitemOpen
  \bibfield  {author} {\bibinfo {author} {\bibfnamefont {D.~I.}\ \bibnamefont
  {Pushkarov}},\ }\href@noop {} {\emph {\bibinfo {title} {Quasiparticle theory
  of defects in solids}}}\ (\bibinfo  {publisher} {World scientific},\ \bibinfo
  {year} {1991})\BibitemShut {NoStop}%
\end{thebibliography}%

\end{document}